\documentclass[twocolumn]{autart}    

\usepackage{graphicx}          
\usepackage{bbding} 
\usepackage{booktabs}
\usepackage{amsmath}
\usepackage{amssymb}
\usepackage{subfigure}
\usepackage{cite}
\usepackage{color}

\begin{document}

\begin{frontmatter}

\title{Stealthy Measurement-Aided Pole-Dynamics Attacks with Nominal Models\thanksref{footnoteinfo}} 

\thanks[footnoteinfo]{This paper was not presented at any IFAC meeting. Corresponding author Changda Zhang. Email: changdazhang@shu.edu.cn.}

\author[SHU]{Dajun Du}\ead{ddj@i.shu.edu.cn},    
\author[SHU]{Changda Zhang}\ead{changdazhang@shu.edu.cn},               
\author[SHU]{Chen Peng}\ead{c.peng@shu.edu.cn},
\author[SHU]{Minrui Fei}\ead{mrfei@staff.shu.edu.cn},  
\author[UOL]{Huiyu Zhou}\ead{hz143@leicester.ac.uk}

\address[SHU]{Shanghai Key Laboratory of
Power Station Automation Technology, School of Mechatronic Engineering
and Automation, Shanghai University, Shanghai 200444, China}  
\address[UOL]{School of Computing and Mathematical Sciences, University of Leicester, Leicester
LE1 7RH, U.K.}             


\begin{abstract}                          
When traditional pole-dynamics attacks (TPDAs) are implemented with nominal models, model mismatch between exact and nominal models often affects their stealthiness, or even makes the stealthiness lost. To solve this problem, our current paper presents a novel stealthy measurement-aided pole-dynamics attacks (MAPDAs) method with model mismatch. Firstly, the limitations of TPDAs using exact models are revealed, where exact models help ensure the stealthiness of TPDAs but model mismatch severely influences its stealthiness. Secondly, to handle model mismatch, the proposed MAPDAs method is designed by using a model reference adaptive control strategy, which can keep the stealthiness. Moreover, it is easier to implement as only the measurements are needed in comparison with the existing methods requiring both the measurements and control inputs. Thirdly, the performance of the proposed MAPDAs method is explored using convergence of multivariate measurements, and MAPDAs with model mismatch have the same stealthiness and similar destructiveness as TPDAs. Specifically, MAPDAs with adaptive gains will remain stealthy at an acceptable detection threshold till destructiveness occurs. Finally, experimental results from a networked inverted pendulum system confirm the feasibility and effectiveness of the proposed method.
\end{abstract}

\end{frontmatter}

\section{Introduction}
Networked control systems (NCSs) \cite{ZhangHan:20}, \cite{ZhangPeng:19}, \cite{ShenPeter:18} deploy communication networks to exchange information between physical entities such as plants, sensors and controllers. Compared with traditional control systems, NCSs eliminate unnecessary wiring, reduce system complexity and cost, and improve system performance. However, the usage of networks makes NCSs open to the outer space and thus be vulnerable to cyber attacks. Recently, there are several attack incidents (\emph{e.g.}, Stuxnet-like attacks on nuclear facilities \cite{TianTan:20} and Blackenergy on power grids \cite{SaxXiong:21}). In this context, it is not surprising that seeking promising solutions to various attacks has attracted wide attention in the community.

The majority of efforts have been made to address a critical question: What degree of attacks can a system bear with the stealthiness before destructiveness is met? Destructiveness means that attacks intentionally drive system state to cross admissible limit, whilst the stealthiness indicates that attacks hide from detectors. There are many types of stealthy attacks, including model-free attacks (\emph{e.g.}, replay attacks \cite{XuLi:21}, optimal linear attacks \cite{GuoShi:17} and switching location attacks \cite{LiuWu:17}) and model-based attacks (\emph{e.g.}, undetectable linear attacks \cite{SongShi:19}, feedback-loop covert attacks \cite{MikZhang:21}, zero-dynamic attacks \cite{TeiShames:15} and pole-dynamics attacks \cite{KimEun:21}). Unlike model-free attacks, model-based attacks rely on a deliberate model to design attacks. A review of recent model-based attacks has been carried out in Section 1.1 of the supplementary materials \cite{DuZhang:22}. When an exact model has been known, model-based attacks can be of the stealthiness whether or not they engage the measurements, control inputs or both.
\begin{table*}[!t]
\centering
\caption{Comparison between the existing construction methods of stealthy attacks and the proposed method.}
\label{Tabsum}
\begin{tabular}{lllllll}
\toprule
  References & Name & Type & Model Required & Type of Plant & MR$^1$ & CIR$^2$\\
\midrule
  \cite{SongShi:19}    & ULAs$^3$        & MBAs$^4$     & Exact model    & Arbitrary     & \XSolidBrush$^{5}$   & \XSolidBrush\\
  \cite{MikZhang:21}   & FLCAs$^{6}$    & MBAs         & Exact model        & Arbitrary         & \XSolidBrush          & \XSolidBrush\\
  \cite{TeiShames:15}    & ZDAs$^{7}$     & MBAs         & Exact model        & Non-minimum phase  & \XSolidBrush          & \XSolidBrush\\
  \cite{KimEun:21}    & PDAs            & MBAs         & Exact model        & Unstable pole-dynamics  & \XSolidBrush          & \XSolidBrush\\
  \cite{LiYang:18}  & DFLCAs$^{8}$   & IMBAs$^{9}$ & No model        & Arbitrary         & \CheckmarkBold$^{10}$ & \CheckmarkBold\\
  \cite{LiXie:19}   & TLCAs$^{11}$    & IMBAs        & No model        & Arbitrary         & \CheckmarkBold        & \CheckmarkBold\\
  \cite{ParkLee:19}   & RZDAs$^{12}$    & IMBAs        & Nominal model & Non-minimum phase         & \CheckmarkBold        & \CheckmarkBold\\
  \cite{JeonEun:19}   & RPDAs$^{13}$    & IMBAs        & Nominal model        & Unstable pole-dynamics         & \CheckmarkBold        & \CheckmarkBold\\
  The proposed method   & MAPDAs           & IMBAs        & Nominal model        & Unstable pole-dynamics         & \CheckmarkBold        & \XSolidBrush\\
\bottomrule
\multicolumn{7}{l}{$^1$Measurement required. $^2$Control input required. $^3$Undetectable linear attacks. $^4$Model-based attacks.}\\
\multicolumn{7}{l}{$^{5}$Not required. $^{6}$Feedback-loop covert attacks. $^{7}$Zero-dynamics attacks.$^{8}$Data-driven FLCA. $^{9}$Improved MBA. }\\
\multicolumn{7}{l}{$^{10}$Required. $^{11}$Data-driven two-loop covert attacks. $^{12}$Robust ZDA. $^{13}$Robust PDA.}\\
\end{tabular}
\end{table*}

However, it is unrealistic to retrieve exact models used by the attacker or defender in many industrial control systems, leading to model mismatch between exact and nominal models. With nominal models, model-based attacks may lose their stealthiness. This brings a consequent question: Are model-based attacks helpless against model mismatch? The answer is no, and there actually are some improved model-based attacks methodologies, \emph{e.g.}, data-driven feedback-loop/two-loop covert attacks \cite{LiYang:18}, \cite{LiXie:19}, robust zero-dynamics attacks \cite{ParkLee:19} and robust PDAs \cite{JeonEun:19}. A review of the existing improved model-based attacks techniques will be carried out in Section 1.1 of the supplementary materials \cite{DuZhang:22}.

Although the existing improved model-based attacks methods have provided promising performance, control inputs are indispensable to the outcome of these methods, \emph{e.g.}, both control inputs and the measurements are required to design the attack mechanisms. This brings a new question: Can improved model-based attacks methods without using control inputs be working against model mismatch between exact and nominal models? In a practical sense, there exist several vulnerable-sensor-network-only NCSs (especially Internet of Things applications \cite{LyuChen:18}) where a vulnerable wireless network may be used to link the sensors and the controller, and reliable cable networks may be used to connect the controller with the plant. Therefore, from the perspective of a defender, the above question is equivalent to this one: Are vulnerable-sensor-network-only NCSs safe enough from these stealthy attacks thanks to model mismatch between exact and nominal models?

Motivated by the above observations, the following challenges and difficulties will be addressed:
\begin{enumerate}
  \item
  Traditional pole-dynamics attacks (TPDAs) are implemented with exact models, which is impractical for some attackers. They have no choice but to use nominal models to design attacks, however model mismatch between exact and nominal models may lead to decline or even loss of stealthiness. Therefore, how to reveal the limitations of TPDAs with nominal models is the first challenge.
  \item
  Some popular techniques (\emph{e.g.}, robust control and data driven) can be employed to improve the stealthiness of TPDAs with nominal models requiring complete and accurate measurements and control inputs. It is difficult for the attacker to launch attacks by obtaining these signals especially in vulnerable-sensor-network-only NCSs. Therefore, how to propose a new attack method without control input is the second challenge.
  \item
  The existing improved model-based attacks (\emph{e.g.}, robust zero-dynamics attacks and robust PDAs) have been mainly designed for single-input-single-output systems, which have rarely been implemented in multiple-input-multiple-output (MIMO) systems. When the above proposed attack method is employed in MIMO systems, identification of stealthiness and destructiveness is the third challenge.
\end{enumerate}

To deal with the above challenges and difficulties, this paper presents a stealthy measurement-aided pole-dynamics attacks (MAPDAs) method with model mismatch for uncertain vulnerable-sensor-network-only NCSs. Comparative analysis between the existing methods and the proposed method is listed in Table~\ref{Tabsum}. The existing methods are mainly based on an exact model or a nominal model but require complete and accurate control inputs, but this paper has revealed the limitations of TPDAs, proposed the new MAPDAs method with model mismatch, and provided the proof of stealthiness and destructiveness of MAPDAs. The main contributions of this paper are summarized as follows:
\begin{enumerate}
  \item
  The limitations of TPDAs using exact models are revealed, where exact models can ensure the stealthiness of TPDAs but model mismatch between the exact and nominal models may cause TPDAs to lose the stealthiness.
  \item
  To handle model mismatch, a new MAPDAs method is proposed using a model reference adaptive control strategy, which can keep the stealthiness. Moreover, it is easier to implement as only the measurements are needed in comparison with the existing methods requiring both the measurements and control inputs.
  \item
  The stealthiness and destructiveness of the proposed MAPDAs in MIMO systems is explored by investigating the convergence of multivariate measurements, where MAPDAs with model mismatch have the same stealthiness and similar destructiveness as TPDAs. Specifically, MAPDAs with adaptive gains will remain stealthy at an acceptable detection threshold till destructiveness occurs.
\end{enumerate}

The reminder of this paper is organized as follows. Section 2 is problem formation, where the limitations of TPDAs with model mismatch are discussed. Section 3 describes the proposed MAPDAs and analyzes the performance. Section 4 provides the experiments where TPDAs and MAPDAs are embedded in the practical networked inverted pendulum visual servo system (NIPVSS), followed by the conclusions made in Section 5.

\begin{rem}
Due to space constraints, some necessary contents are placed in the supplementary materials \cite{DuZhang:22}.
\end{rem}

\textbf{Notation.} For a matrix $P$, $P>0$ denotes that $P$ is positive definite symmetric. The one vector (all elements are 1) is denoted by $\textbf{1}_{n} \in \mathbb{R}^{n}$. Table~A.1 in Section 1.2 of supplementary materials \cite{DuZhang:22} summarizes the notations most frequently used throughout the rest of the paper.

\section{Problem Formulation}
\subsection{NCSs under TPDAs with Exact Model}
\begin{figure}[!t]
  \centering
  \includegraphics[width=0.47\textwidth]{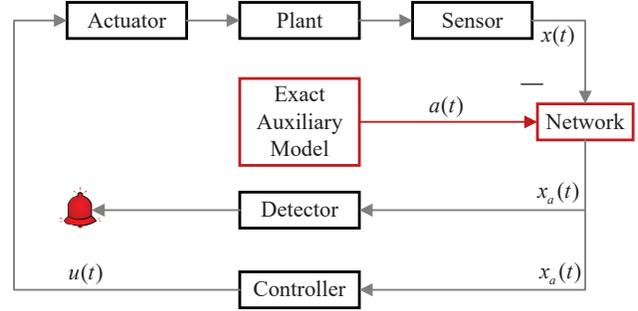}
  \caption{Framework of NCSs for TPDAs with exact auxiliary models.}
  \label{fig1}
\end{figure}

The framework of NCSs for TPDAs with exact auxiliary model ({\em i.e.}, exact model) is shown in Fig.~\ref{fig1}. Firstly, the sensor obtains the measurement $x(t)$ from the plant. Then, $x(t)$ will be transmitted to the estimator and controller via networks, becoming $x_a(t)$ due to injection of attack signals $a(t)$ from possible TPDAs with exact auxiliary model. Using $x_a(t)$, the controller calculates control input $u(t)$ that is sent to the actuator to stabilize the plant and the estimator judges whether or not there exists an attack, and if there is an attack, the alarm will be triggered.

\begin{rem}
Fig.~\ref{fig1} shows the framework of NCSs for TPDAs with exact auxiliary model, where the construction of TPDAs adopts exact auxiliary model (\ref{eq3A3}) that only needs the matrix $A$ of physical system (\ref{eq3A1}) and does not use $x(t)$, see \cite{KimEun:21}, \cite{JeonEun:19}. However, when there exists model mismatch between exact auxiliary model and nominal model (i.e., the attackers can not obtain exact auxiliary model), $x(t)$ and $u(t)$ have been applied to constructing model-based attacks for the stealthiness. Thus, $u(t)$ needs to be connected into exact auxiliary model box in this case, see \cite{LiYang:18}, \cite{LiXie:19}, \cite{ParkLee:19}, \cite{JeonEun:19}.
\end{rem}

Consider continuous linear time-invariant (LTI) plant\footnote{Strictly speaking, the poles set of an LTI system is the subset of the eigenvalues set of $A$ \cite[Lec. 19.2]{Hesp:2006}. For this reason, $\dot x(t)=Ax(t)$ is called pole-dynamics.}
\begin{align}
  \dot x(t) &= Ax(t) + Bu(t), \label{eq3A1}\hfill \\
  z(t) &= Cx(t), \label{eq3A2} \hfill
\end{align}
where $x(t) \in \mathbb{R}^{p}$ is system state and the measurement, $u(t) \in \mathbb{R}^{m}$ is control input, $z(t) \in \mathbb{R}^{q}$ is  controlled output, and $A$, $B$, and $C$ are constant matrices with appropriate dimensions. Without loss of generality, it is considered that (1) is controllable.

System state $x(t)$ will be transmitted to the controller of TPDAs via networks. TPDAs can maintain a continuous exact auxiliary model ${\mathcal{A}_{c,t}}(A)$ \cite{KimEun:21}, \cite{JeonEun:19}:
\begin{subequations}
\label{eq3A3}
\begin{align}
\label{eq3A3a} {{\dot x}_{\rm eam}}(t) &= A{x_{\rm eam}}(t), \hfill \\
\label{eq3A3b} a(t) &= {x_{\rm eam}}(t), \hfill
\end{align}
\end{subequations}
where $x_{\rm eam}(t)$ and $a(t)$ are the state and the output of $\mathcal{A}_{c,t}(A)$. In a network, $a(t)$ may be subtracted from $x(t)$, and thus the network output becomes
\begin{equation}
x_a(t)=x(t)-a(t).
\label{eq3A4}
\end{equation}
Using $x_a(t)$, the controller calculates the control input
\begin{equation}
u(t)=K x_a(t),
\label{eq3A5}
\end{equation}
where $K$ is the controller gain and has been designed to make $\Phi:=A+BK$ stable (\emph{i.e.}, the eigenvalues of $\Phi$ are located on the closed left half-plane). Then, $u(t)$ will be sent to the actuator for stabilizing the plant.

To detect attacks, the common norm-based test is performed by the detector, \emph{i.e.}, if there is
\begin{equation}
\left\| {{x_a}(t)} \right\| < \epsilon,
\label{eq3A6}
\end{equation}
where $\epsilon >0$ is a user-defined threshold, it means that there is no attack, otherwise, attacks emerge.
\begin{rem}
The threshold $\epsilon$ of the detector (\ref{eq3A6}) is the key to examine the validity of attack detection. The existing threshold selection methods generally include statistical analysis \cite{HeydtGraf:10}, theoretical derivation \cite{MoSinopoli:09}, machine learning \cite{ZhaoZhong:20}, etc. To determine the proper threshold, the method of statistical analysis is used in Section 4.
\end{rem}

To analyse the performance of attacks, the definitions of stealthiness and destructiveness in time period $\mathcal{T}:=[t_0,t_f]$ of attacks are given in the following, where $t_0$, $t_f$ are initial and finishing instants of attacks, respectively.
\setcounter{thm}{0}
\begin{defn}[$\epsilon$-stealthiness]
(cf. \cite{KungDey:17}) An attack is said to be with $\epsilon$-stealthiness on the detector in $\mathcal{T}$ when (\ref{eq3A6}) for $t \in \mathcal{T}$ always holds.
\end{defn}
\begin{defn}[$\xi$-destructiveness]
(cf. \cite{ParkLee:19}) An attack is said to be with $\xi$-destructiveness on the controlled output $z(t_f)$ if
\begin{equation}
\left\| z(t_f) \right\| \geqslant  \xi,
\label{eq3A7}
\end{equation}
where $\xi$ is the admissible state limit. Specifically, $\left\| z(t) \right\| < \xi $ for $t \in \mathcal{T}$ will run under control and $\left\| z(t) \right\| \geqslant  \xi $ for $t \in \mathcal{T}$ is actively out of control (e.g., takes active protection measures) to avoid possible severe accidents.
\end{defn}
It is well believed that an ideal attack in $\mathcal{T}$ should be with both $\xi$-destructiveness and $\epsilon$-stealthiness. We may witness a more dangerous scenario where a quasi-ideal attack is with $\epsilon$-stealthiness and with no $\xi$-destructiveness in $\mathcal{T}$, but the controlled output is driven very close to $\xi$. A quasi-ideal attack could be on the synchronous machines \cite{EndPil:11}, where the attack will not make rotation rates of synchronous machines cross the admissible limit, but it pushes the rotation rate to be high. This will remarkably shorten the life of synchronous machines and even cause accidents. For simplicity, our current paper only focuses on ideal attacks.

\subsection{Performance of TPDAs with Exact Auxiliary Models}
Based on the above NCSs under TPDAs, the definitions and impact analysis of TPDAs with $\mathcal{A}_{c,t}(A)$ (\ref{eq3A3}) in \cite{JeonEun:19}, the stealthiness and destructiveness of TPDAs with $\mathcal{A}_{c,t}(A)$ (\ref{eq3A3}) are presented in the following Theorem~\ref{T1}.
\setcounter{thm}{0}
\begin{thm}
\label{T1}
Considering system (\ref{eq3A1})-(\ref{eq3A5}) under TPDAs with $\mathcal{A}_{c,t}(A)$ (\ref{eq3A3}), if $\Phi$ is stable, $A$ is unstable (i.e., at least one of eigenvalues of $A$ is located on the open right half-plane), and $x_{eam}(t_0)$ does not satisfy the item (i) or (ii) of Lemma A.1 in Section 2.1 of the supplementary materials \cite{DuZhang:22}, then
\begin{equation}
\label{eq3B1}
\exists \epsilon,\left\| {{x_a}(t)} \right\| < \epsilon,t \in [{t_0},\infty ).
\end{equation}
The norm of system state becomes unbounded, i.e.,
\begin{equation}
\mathop {\lim }\limits_{t \to \infty } \left\|x(t)\right\| \to \infty.
\label{eq3B2}
\end{equation}
\end{thm}
\begin{pf}
The proof is given in Section 2.2 of the supplementary materials \cite{DuZhang:22}.
\end{pf}
\setcounter{thm}{3}
\begin{rem}
For Theorem 1, there possibly exist two cases for TPDAs with $\mathcal{A}_{c,t}(A)$ (\ref{eq3A3}), i.e., $\epsilon > {\sup}_{\rm eam} \left\| {x_a}\right\|$ (${\sup}_{\rm eam} \left\| {x_a} \right\|$ represents the upper bound of $\left\|x_a(t)\right\|$ under TPDAs with $\mathcal{A}_{c,t}(A)$ (\ref{eq3A3})) and $\epsilon \leqslant {\sup}_{\rm eam} \left\| {x_a}\right\|$, which is shown in Fig.~A.1(a) of Section 2.3 in the supplementary materials \cite{DuZhang:22}. Therefore, for a given $\epsilon$, when a small enough ${\sup}_{\rm eam} \left\| {x_a}\right\|$ is selected, TPDAs with $\mathcal{A}_{c,t}(A)$ (\ref{eq3A3}) can be with $\xi$-destructiveness and $\epsilon$-stealthiness in $\mathcal{T}$.
\end{rem}
\begin{rem}
For Lemma A.1 in Section 2.1 of the supplementary materials \cite{DuZhang:22}, we examine whether or not the initial value of ${x_{{\rm eam},i}}({t_0})$ [i.e., $x_{{\rm eam},i}(t_0)$ corresponding to the $i^{th}$ eigenvalue $\lambda_{i}$ of the matrix $A$, $i=1,\ldots,p$] equals to zero. There are two cases for $x(t)$: (1) If $x_{\rm eam}(t_0)$ satisfies the initial condition in Lemma A.1, $\mathop {\lim }\limits_{t \to \infty } {x_{{\text{eam}}}}(t) = 0$ even if $A$ is unstable. Furthermore, according to (\ref{eq3A3}), (\ref{eq3A4}) and (A.3) in Section 2.2 of the supplementary materials \cite{DuZhang:22}, $\mathop {\lim }\limits_{t \to \infty } x(t) = 0$ so that (\ref{eq3B2}) will not hold; (2) If $x_{\rm eam}(t_0)$ does not satisfy the initial condition in Lemma A.1, $\mathop {\lim }\limits_{t \to \infty } \left\| {{x_{\rm eam}}(t)} \right\| \to \infty $ as $A$ is unstable. Furthermore, according to (\ref{eq3A3}), (\ref{eq3A4}) and (A.3), $\mathop {\lim }\limits_{t \to \infty } \left\| {x(t)} \right\| \to \infty $, and (\ref{eq3B2}) in Theorem~\ref{T1} holds.
\end{rem}

\subsection{Limitation of TPDAs with Nominal Models}
Although the above stealthiness and destructiveness of TPDAs with $\mathcal{A}_{c,t}(A)$ (\ref{eq3A3}) look promising, it is unrealistic to obtain the exact model for the attacker (even for the defender). When the attacker only knows the nominal model of uncertain NCSs (\emph{i.e.}, the nominal model $A_n$ of $A$), they have to perform TPDAs with a continuous nominal auxiliary model ${\mathcal{A}_{c,t}}({A_n})$
\begin{subequations}\label{eq3C1}
\begin{align}
\label{eq3C1a}  {{\dot x}_{{\text{nam}}}}(t) &= {A_n}{x_{{\text{nam}}}}(t), \hfill \\
\label{eq3C1b}  a(t) &= {x_{{\text{nam}}}}(t), \hfill
\end{align}
\end{subequations}

According to impact analysis of TPDAs with $\mathcal{A}_{c,t}(A)$ (\ref{eq3A3}) in \cite{JeonEun:19}, the limitation of TPDAs with $\mathcal{A}_{c,t}(A_n)$ (\ref{eq3C1}) is presented in the following Theorem~\ref{T2}.
\setcounter{thm}{1}
\begin{thm}
\label{T2}
Considering the system (\ref{eq3A1}), (\ref{eq3A2}), (\ref{eq3A4}), (\ref{eq3A5}) under TPDAs with $\mathcal{A}_{c,t}(A_n)$ (\ref{eq3C1}), if $\Phi$ is stable, $A_n$ is unstable, and $x_{\rm nam}(t_0)$ does not satisfy the item (i) or (ii) of Lemma A.2 in Section 2.5 of the supplementary materials \cite{DuZhang:22}, then
\begin{equation}
\mathop {\lim }\limits_{t \to \infty } \left\|x_a(t)\right\| \to \infty.
\label{eq3C2}
\end{equation}
The norm of system state becomes unbounded, i.e., (\ref{eq3B2}).
\end{thm}
\begin{pf}
The proof is given in Section 2.4 of the supplementary materials \cite{DuZhang:22}.
\end{pf}
\setcounter{thm}{5}
\begin{rem}
For Theorem 2, there possibly exist two case for TPDAs with $\mathcal{A}_{c,t}(A_n)$ (\ref{eq3C1}), i.e., $\epsilon  > \left\| {x_a(t_f)} \right\|$ and $\epsilon \leqslant \left\| {x_a(t_f)} \right\|$ ($t_f$ is finishing instant of attacks), which is shown in Fig.~A.1(b) of Section 2.3 in the supplementary materials \cite{DuZhang:22}. Therefore, when a small $\epsilon$ is selected, TPDAs with $\mathcal{A}_{c,t}(A_n)$ (\ref{eq3C1}) are with $\xi$-destructiveness but with no $\epsilon$-stealthiness in $\mathcal{T}$.
\end{rem}
\begin{rem}
For Lemma A.2 in Section 2.5 of the supplementary materials \cite{DuZhang:22}, we have analysed whether or not the initial value of $x_{{\rm nam},i}(t_0)$ (i.e., $x_{{\rm nam},i}(t_0)$ corresponding to the $i^{th}$ eigenvalue $\lambda_{n,i}$ of the matrix $A_n$) equals to zero. There are two cases for $x_a(t)$: (1) If $x_{\rm nam}(t_0)$ satisfies the initial condition in Lemma A.2, $\mathop {\lim }\limits_{t \to \infty } {x_{\rm nam}}(t) = 0$ even if $A_n$ is unstable. Furthermore, according to (\ref{eq3C1}), (\ref{eq3A4}) and (A.5) in Section 2.4 of the supplementary materials \cite{DuZhang:22}, $\mathop {\lim }\limits_{t \to \infty }  {{x_a}(t)} =0 $ so that (\ref{eq3C2}) will not hold; (2) If $x_{\rm nam}(t_0)$ does not satisfy the initial condition in Lemma A.2, $\mathop {\lim }\limits_{t \to \infty } \left\| {{x_{\rm nam}}(t)} \right\| \to \infty $ because $A_n$ is unstable. Furthermore, according to (\ref{eq3C1}), (\ref{eq3A4}) and (A.5), $\mathop {\lim }\limits_{t \to \infty } \left\| {{x_a}(t)} \right\| \to \infty $, and (\ref{eq3C2}) in Theorem~\ref{T2} holds.
\end{rem}
Up to now, we understand the limitations of TPDAs with nominal models, \emph{i.e.}, TPDAs will be with $\xi$-destructiveness but with no $\epsilon$-stealthiness in $\mathcal{T}$. In the next section, to cope with this problem, we will present a measurements and adaptive control based method into the attacks.

\section{Measurement-Aided Pole-Dynamics Attacks}
We have analysed NCSs under TPDAs and the limitations of TPDAs with nominal models in the previosu sections. To solve the problem, a stealthy MAPDAs method using measurements and an adaptive auxiliary model will be designed and discussed.

\subsection{Design of MAPDAs}
\begin{figure}[!t]
  \centering
  \subfigure[]{\includegraphics[width=0.47\textwidth]{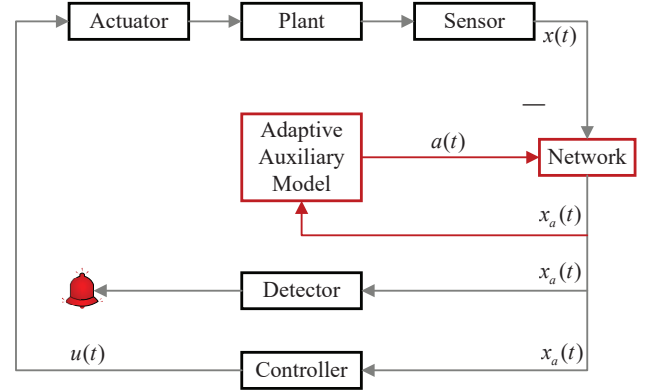}} \\
  \subfigure[]{\includegraphics[width=0.47\textwidth]{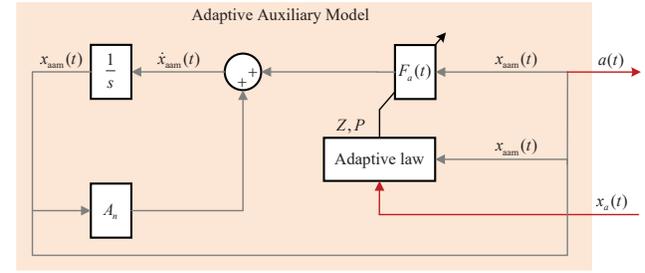}}
  \caption{(a) Framework of NCSs under MAPDAs with an adaptive auxiliary model. (b) Framework of an adaptive auxiliary model.}
  \label{fig2}
\end{figure}
The framework of NCSs under MAPDAs with an adaptive auxiliary model is shown in Fig.~\ref{fig2}(a) and the framework of adaptive auxiliary model is shown in Fig.~\ref{fig2}(b). Firstly, the sensor collects $x(t)$ from the plant. Then, $x(t)$ will be transmitted to the detector and controller via network, becoming $x_a(t)$ due to possible attacks. The attacker obtains $x_a(t)$ and uses it to construct adaptive auxiliary model with state $x_{\rm aam}(t)$ and output $a(t)$, where $x_a(t)$ and $x_{\rm aam}(t)$ are used to produce adaptive gain $F_a(t)$ based on the designed adaptive laws, and $F_a(t)$ is used to update $x_{\rm aam}(t)$ and $a(t)$. Using $x_a(t)$, the controller calculates $u(t)$ that is sent to the actuator to stabilize the plant and the estimator scrutinizes whether or not there is an attack, and if there is an attack, the alarm will be triggered.

Consider continuous LTI plant (\ref{eq3A1}) and (\ref{eq3A2}) and that the attacker has nominal models (\emph{i.e.}, $A_n$, $B_n$, $K_n$ of NCSs) and can obtain the data in the network. Motivated by the direct model reference adaptive control \cite{Tao:14}, \cite{KerBuss:17} (that is actually is simplified in this paper, and specifically the external command is zero and the reference model specifies the desired response with zero), they can perform MAPDAs with a continuous adaptive auxiliary model ${\mathcal{A}_{a,t}}({A_n},\Phi,x_a)$:
\begin{subequations}\label{eq4A1}
\begin{align}
\label{eq4A1a} {\dot x_{\rm aam}}(t) &= \left({A_n}+F_a(t)\right){x_{\rm aam}}(t), \hfill \\
\label{eq4A1b} \dot F_a(t) &=  Z P{x_a}(t)x_{\rm aam}^T(t), \hfill\\
\label{eq4A1c} a(t) &= {x_{{\text{aam}}}}(t), \hfill \\
\label{eq4A1d} - Q &= {\Phi ^T}P + P\Phi, \hfill
\end{align}
\end{subequations}
where $x_{\rm aam}(t)$ and $a(t)$ are respectively state and output of $\mathcal{A}_{a,t}(A_n,\Phi,x,x_a)$, $F_a(t)$ is time-varying adaptive gain, $Z>0$, $P>0$, $Q>0$ are constant matrices, and $x_a(t)$ is the network output in (\ref{eq3A4}). The control input is $u(t)$ in (\ref{eq3A5}) and the detector with the test (\ref{eq3A6}) is used to detect attacks.
\setcounter{thm}{7}
\begin{rem}
The goal of MAPDAs with ${\mathcal{A}_{a,t}}({A_n},\Phi,x_a)$ (\ref{eq4A1}) is to drive $x_a(t)$ to converge to 0 as that of TPDAs with ${\mathcal{A}_{c,t}}(A)$ (\ref{eq3A3}). To achieve this goal, different from TPDAs with ${\mathcal{A}_{c,t}}(A_n)$ (\ref{eq3C1}), additional adaptive gain $F_a(t)$ and the measurement $x_a(t)$ are required for MAPDAs with ${\mathcal{A}_{a,t}}({A_n},\Phi,x_a)$ (\ref{eq4A1}). By using MAPDAs with ${\mathcal{A}_{a,t}}({A_n},\Phi,x_a)$ (\ref{eq4A1}) and considering the system (\ref{eq3A1}), (\ref{eq3A2}), (\ref{eq3A4}), (\ref{eq3A5}), the dynamics of $x_a(t)$ becomes
\begin{align}
{{\dot x}_a}(t) &= \Phi {x_a}(t)+ \left( {A -{F_a}(t) - {A_n}} \right){x_{\rm aam}}(t). \hfill
\label{eq4A2}
\end{align}
In (\ref{eq4A2}), $x_a(t)$ will be driven to asymptotically converge to 0, which is proved by using Lyapunov stability theory in the next subsection.
\end{rem}

\subsection{Performance of MAPDAs}
MAPDAs with ${\mathcal{A}_{a,t}}({A_n},\Phi,x_a)$ (\ref{eq4A1}) have been designed above and its performance will be presented in the following Theorem~\ref{T3}.
\setcounter{thm}{2}
\begin{thm}
\label{T3}
Considering the systems (\ref{eq3A1}), (\ref{eq3A2}), (\ref{eq3A4}), (\ref{eq3A5}) under MAPDAs with ${\mathcal{A}_{a,t}}({A_n},\Phi,x_a)$ (\ref{eq4A1}), for $Q>0$ and $Z>0$, if $\Phi$ is stable, then
\begin{equation}
\mathop {\lim }\limits_{t \to \infty } {x_a}(t) = 0.
\label{eq4B1}
\end{equation}
\end{thm}
\begin{pf}
The proof is given in Section 3.1 of the supplementary materials \cite{DuZhang:22}.
\end{pf}
\setcounter{thm}{8}
\begin{rem}
For Theorem 3, there possibly exist three types of MAPDAs, i.e., climbing type ($\epsilon < {\sup _{\rm aam}}\left\| {{x_a}} \right\|$, ${\sup _{\rm aam}}\left\| {{x_a}} \right\|$ represents the upper bound of $\left\| x_a(t) \right\|$ under the proposed MAPDAs), peak type ($\epsilon = {\sup _{\rm aam}}\left\| {{x_a}} \right\|$) and descending type ($\epsilon > {\sup _{\rm aam}}\left\| {{x_a}} \right\|$), which is shown in Fig. A.2 of Section 3.2 in the supplementary materials \cite{DuZhang:22}. It can provide the guideline for the attacker and defender. For the view of the attacker, they can select the proper parameters of MAPDAs for small $\left\| {{x_a}} \right\|$ (good stealthiness). However, for the view of the defender, it is suggested not to select too big threshold $\epsilon$. This paper mainly focuses on the new stealthy MAPDAs method from the perspective of the attacker, so the discussion of these three types is valuable to help to select the proper parameters of MAPDAs.
\end{rem}
\begin{rem}
The selection of parameters $Q$ and $Z$ of the proposed MAPDAs with ${\mathcal{A}_{a,t}}({A_n},\Phi,x_a)$ (\ref{eq4A1}) will affect the dynamics of $x_a(t)$ and $x(t)$, i.e., the upper bound of $\left\| x_a(t)\right\|$ and the limit-crossing speed of $\left\| x(t)\right\|$. When $Q$ and $Z$ are selected improperly, the upper bound of $\left\| x_a(t)\right\|$ could be close to the threshold and with a high limit-crossing speed of $\left\| x(t)\right\|$. On the contrary, when $Q$ and $Z$ are selected properly, the upper bound of $\left\| x_a(t)\right\|$ could be far less than the threshold and with a low limit-crossing speed of $\left\| x(t)\right\|$. These two cases are shown in Section 4. The parameters  $Q$ and $Z$ can be selected by using some popular methods such as trial-and-error method, optimization algorithm and so on.
\end{rem}
However, the attackers cannot obtain $\Phi$ and thus they cannot calculate $P$ by (\ref{eq4A1d}) and $F_a(t)$ in (\ref{eq4A1b}), making MAPDAs with ${\mathcal{A}_{a,t}}({A_n},\Phi,{x_a})$ (\ref{eq4A1}) unable to operate. To cope with this problem, the ideal ${\mathcal{A}_{a,t}}({A_n},\Phi,{x_a})$ (\ref{eq4A1}) is slightly regulated into ${\mathcal{A}_{a,t}}({A_n},\Phi_n,{x_a})$
\begin{equation}
\begin{gathered}
  \text{(\ref{eq4A1a})-(\ref{eq4A1c}),\ and}  \hfill \\
   - Q = \Phi _n^TP + P{\Phi _n}, \hfill \\
\end{gathered}
\label{eq4B2}
\end{equation}
where $\Phi_n := A_n +B_nK_n$ is the nominal part of $\Phi$.
\setcounter{thm}{0}
\begin{cor}
\label{C1}
Considering the systems (\ref{eq3A1}), (\ref{eq3A2}), (\ref{eq3A4}), (\ref{eq3A5}) under MAPDAs with ${\mathcal{A}_{a,t}}({A_n},\Phi_n,x,x_a)$ (\ref{eq4B2}), for $Q>0$ and $Z>0$, if $\Phi$ is stable and (\ref{eq4A1d}) holds, then (\ref{eq4B1}) will hold.
\end{cor}
\begin{pf}
The proof is similar as that of Theorem 3, which is thus omitted.
\end{pf}
\setcounter{thm}{10}
\begin{rem}
Corollary 1 indicates that after $Q$ has been selected, $P$ can be calculated by using (\ref{eq4B2}). If the selected $Q$ and the calculated $P$ from (\ref{eq4B2}) satisfies (\ref{eq4A1d}), then (\ref{eq4B1}) will hold, i.e., the stealthiness of MAPDAs is achieved. However, it is not easy to obtain exact auxiliary model for the attacker, so (\ref{eq4A1d}) cannot be verified. Therefore, $Q$ needs to be selected by using some methods as discussed in the above Remark 10.
\end{rem}
The problem of PDAs without control inputs against model mismatch is completely solved along the following line $(\ref{eq3A3}) \to (\ref{eq3C1}) \to (\ref{eq4A1}) \to (\ref{eq4B2})$. Firstly, in spite of the promising performance given in Theorem~\ref{T1}, TPDAs with (\ref{eq3A3}) are denied due to model mismatch. Secondly, the limitation of TPDAs with (\ref{eq3C1}) against model mismatch is revealed in Theorem~\ref{T2} from perspective of $\epsilon$-stealthiness and $\xi$-destructiveness. Then, to deal with the limitation, MAPDAs with (\ref{eq4A1}) are designed by introducing the measurements and an adaptive control method, whose performance is given in Theorem~\ref{T3}. Finally, to further cast off the dependence on exact models, the regulated MAPDAs with (\ref{eq4B2}) are developed. The experimental demonstration will be given in the next section.
\begin{rem}
When the open-loop dynamic of the nominal system is stable (i.e., $A_n$ is stable), the proposed MAPDAs cannot ensure that system state is divergent, see the analysis in Section 3.3 of the supplementary materials \cite{DuZhang:22}.
\end{rem}
\begin{rem}
When the considered NCSs is a digital system (e.g., both the sensors and controller are digitized with the sampling periods), the attacker can adopt discrete-time TPDAs (A.12) or (A.13) and MAPDAs (A.14) (see Section 3.4 of the supplementary materials \cite{DuZhang:22}) transformed from continuous TPDAs (\ref{eq3A3}) or (\ref{eq3C1}) and MAPDAs (\ref{eq4A1}). The next aim is to analyse the effectiveness of discrete-time MAPDAs (A.14) (taken as an example) for the digital system. Considering that when the sensors and controller are digitized, the digital system becomes a sampled-data-based hybrid system. For this kind of hybrid system, referring to \cite{ZhangHan:17}, \cite{LingKrava:19}, it is commonly expressed as a time-delay system and stability criterions can be given. Therefore, a time-delay system is given, and its stability criterion on delay-induced continuous MAPDAs (A.16) to guarantee the stealthiness (\ref{eq4B1}) has been proved by Theorem A.1 in Section 3.4 of the supplementary materials \cite{DuZhang:22}. According to (A.16) in Theorem A.1, a delay-induced discrete-time MAPDAs (A.22) is obtained. Note that there exists only one difference between discrete-time MAPDAs (A.14) and delay-induced discrete-time MPADAs (A.22), i.e., the last three items related to the square of the sampling period are additional in (A.22b). When the sampling period is small, discrete-time MAPDAs (A.14) is a proper approximation of delay-induced discrete-time MPADAs (A.22). Since discrete-time MPADAs (A.22) can guarantee the stealthiness (14), discrete-time MAPDAs (A.14) will be effective on guaranteeing the stealthiness (14) for the digital system.
\end{rem}

\section{Experimental Results and Discussion}
\begin{figure}[!t]
  \centering
  \includegraphics[width=0.47\textwidth]{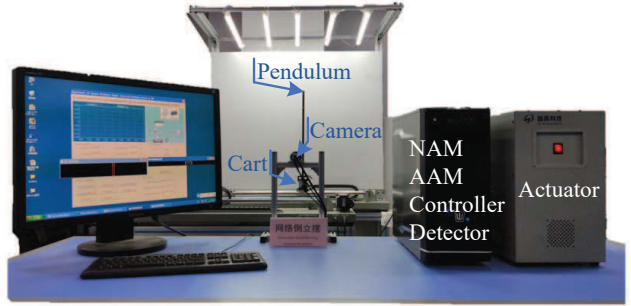}
  \caption{Experimental platform of NIPVSS. NAM: Nominal auxiliary model. AAM: Adaptive auxiliary model.}
  \label{fig4}
\end{figure}
To validate the proposed MAPDAs, we consider the scenario when TPDAs \cite{JeonEun:19}, MAPDAs, DFLCAs \cite{LiYang:18}, and TLCAs \cite{LiXie:19} are embedded in a networked inverted pendulum visual servo system (NIPVSS) \cite{DuZhang:20} in Fig.~\ref{fig4}.

\subsection{Parameters of NIPVSS}
The state of NIPVSS is set as $x(t) = [\alpha (t),\theta (t),\dot \alpha (t),\dot \theta (t)]$, where $\alpha(t)$ is the cart position, $\theta(t)$ is the pendulum angle, $\dot \alpha(t)$ and $\dot \theta(t)$ are the cart and angular velocity, respectively. The acceleration-as-control-input nonlinear differential equation of the inverted pendulum is
\begin{equation}
lmu\cos \theta  + lmg\sin \theta  = J\ddot \theta,
\label{eq5A1}
\end{equation}
where $l$ is the length from the pivot to the center of the pendulum, $m$ is the mass of the pendulum, $g$ is the acceleration of the gravity, $J$ is the moment of the inertia about the pivot of the pendulum, the values of $l$, $m$, $g$, $J$ can be found in \cite{DuZhang:20}, and $u=\ddot \alpha$ is the control input. By linearizing (\ref{eq5A1}) in $\left| {\theta}\right| \leqslant 0.2 rad$ (\emph{i.e.}, $\cos \theta \approx 1$ and $\sin \theta \approx\theta$ in $\left| {\theta}\right| \leqslant 0.2 rad$), the nominal model $A_n$ and $B_n$ of (\ref{eq5A1}) is given by
\[{A_n} = \left[ {\begin{array}{*{20}{c}}
  0&0&1&0 \\
  0&0&0&1 \\
  0&0&0&0 \\
  0&29.4311&0&0
\end{array}} \right],{B_n} = \left[ {\begin{array}{*{20}{c}}
  0 \\
  0 \\
  1 \\
  3.0001
\end{array}} \right].\]
Based on $A_n$, $B_n$ and using $H_{\infty}$ control \cite{DuZhang:20}, the controller is designed as
\[K=K_n=[3.7569,-29.6225,4.0648,-5.4563].\]
The controlled outputs are $\alpha (t)$ and $\theta (t)$. The admissible limits are $\left|\alpha (t)\right|<0.3 m$ and $\left|\theta (t)\right|<0.8 rad$.

\subsection{Threshold of the Detector}
\begin{figure}[!t]
\centering
\includegraphics[width=0.475\textwidth]{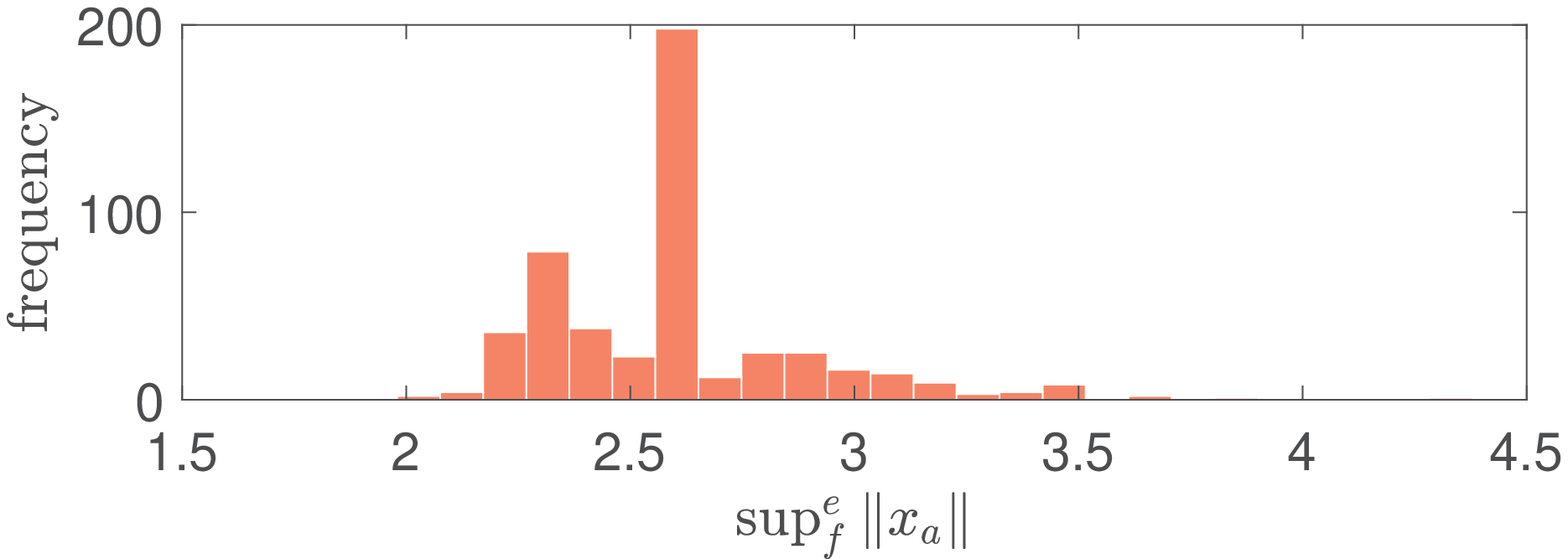}
\caption{The frequency of different ${{{\sup }^e_f}\left\| {{x_a}} \right\|}$ from all 500 experiments.}
\label{fig5}
\end{figure}
To properly set the threshold $\epsilon$ of the detector, in terms of statistical analysis method \cite{HeydtGraf:10}, 500 experiments of attack-free NIPVSS are operated (see Fig.~A.3 and Table A.2 in Section 4.1 of the supplementary materials \cite{DuZhang:22}). The frequency of different ${{{\sup }^e_f}\left\| {{x_a}} \right\|}$ (i.e., the upper bound of attack-free $\left\| x_a(t)\right\|$) are shown in Fig.~\ref{fig5}. It can be seen from Fig.~A.3, Table~A.2 and Fig.~\ref{fig5} that after the state of NIPVSS is stable, the threshold can be set as $\epsilon = 3.1$ based on the 3$\sigma$ principle.

\subsection{Performance of TPDAs and MAPDAs}
For comparison, we construct two types of attacks with the values of $A_n$, $B_n$ and $K_n$. One is TPDAs \cite{JeonEun:19} with $\mathcal{A}_{c,t}(A_n)$ (\ref{eq3C1}) when $t\geqslant 0s$ (\emph{i.e.}, $k \geqslant 0$). Another is the proposed MAPDAs with ${\mathcal{A}_{a,t}}({A_n},\Phi_n,x_a)$ (\ref{eq4B2}) when $t\geqslant 0s$ (\emph{i.e.}, $k \geqslant 0$), and their parameters are set as $Q=I$ and $Z=10000I$ in (\ref{eq4B2}), and $P$ is calculated by using (\ref{eq4B2}) as
\[P = \left[ {\begin{array}{*{20}{c}}
  {1.7760}&{ - 2.0855}&{0.8362}&{ - 0.3231} \\
   * &{10.6948}&{ - 2.9413}&{1.4742} \\
   * & * &{1.0652}&{ - 0.4646} \\
   * & * & * &{0.2755}
\end{array}} \right].\]
The initial condition is set to $x_{\rm nam}(0)=0.0001 \textbf{1}$ in (\ref{eq3C1}), and $x_{\rm aam}(0)=0.0001 \textbf{1}$, $F_a(0)=I$ in (\ref{eq4B2}). According to the condition of Theorem 2, it is verified in Section 4.2.1 of the supplementary materials \cite{DuZhang:22} that $x_{\rm nam}(0)=0.0001 \textbf{1}$ does not satisfy the item (i) or (ii) of Lemma A.2 in Section 2.5 of the supplementary materials \cite{DuZhang:22}. Therefore, it satisfies the initial condition of Theorem 2.
\begin{figure}[!t]
  \centering
  \subfigure[$\alpha(t)$]{\includegraphics[width=0.47\textwidth]{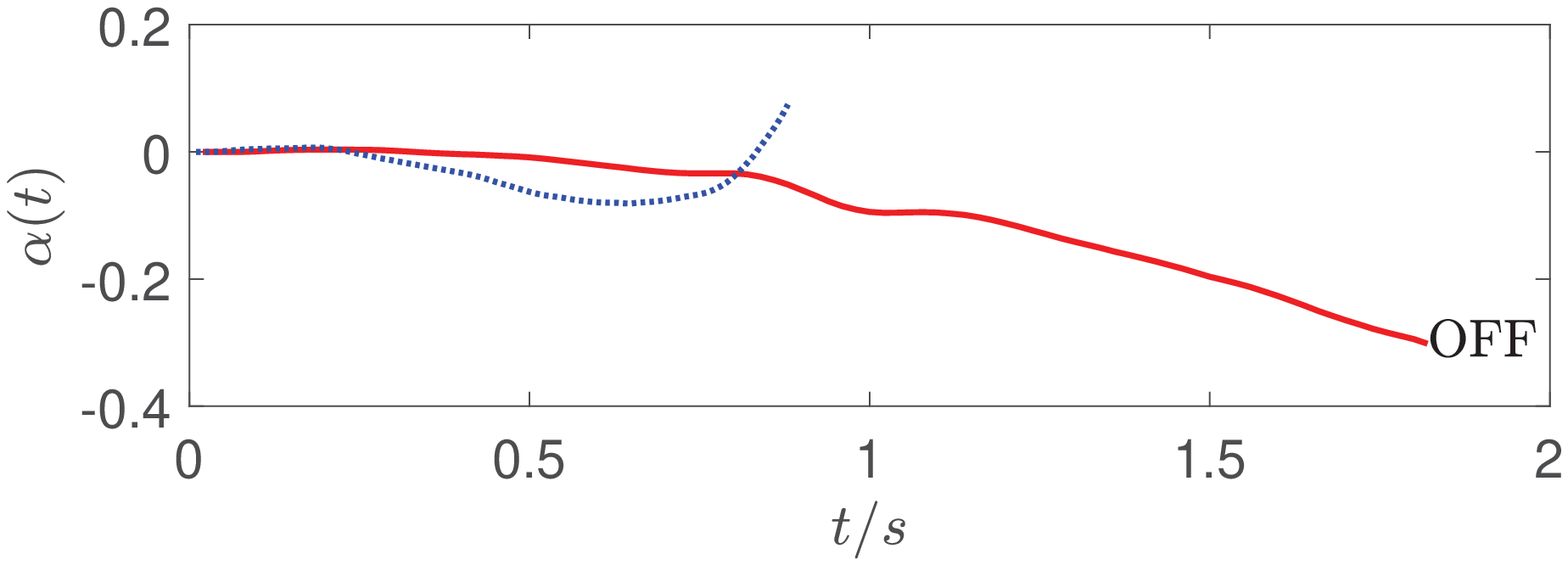}}\\ \vspace{-0.15in}
  \subfigure[$\theta(t)$]{\includegraphics[width=0.47\textwidth]{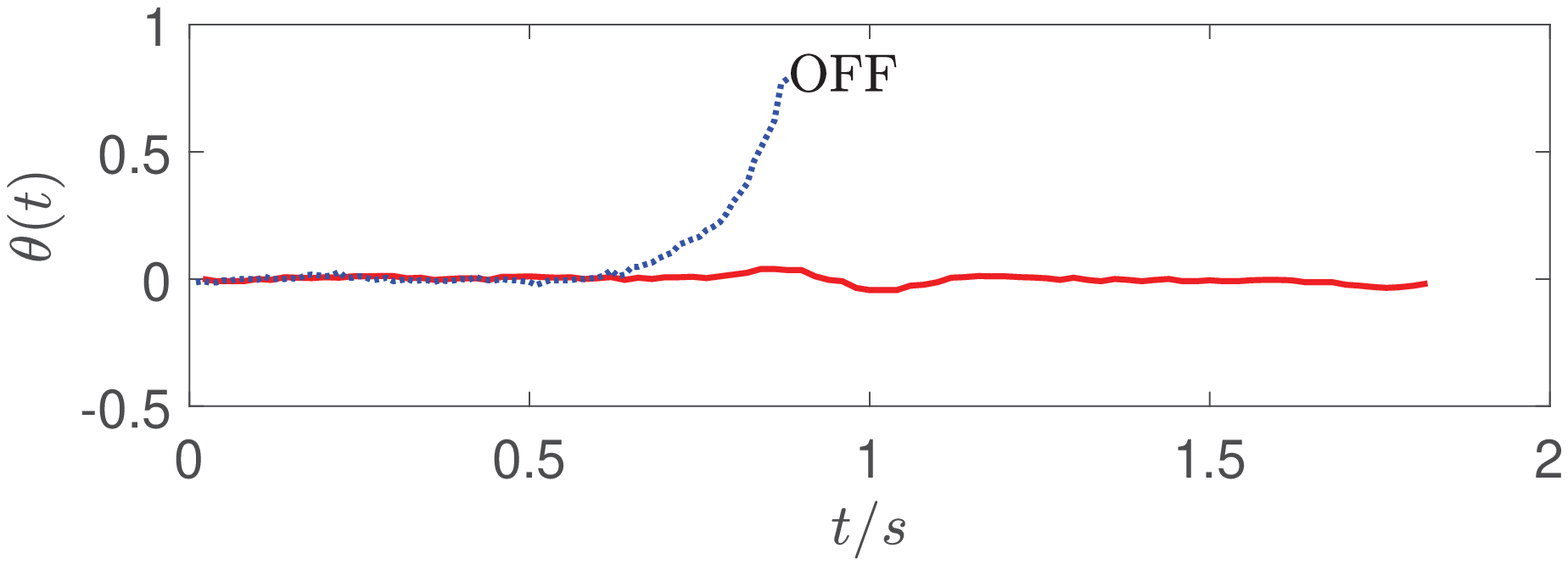}} \\ \vspace{-0.15in}
  \subfigure[$\left\| {x_a}(t) \right\|$]{\includegraphics[width=0.47\textwidth]{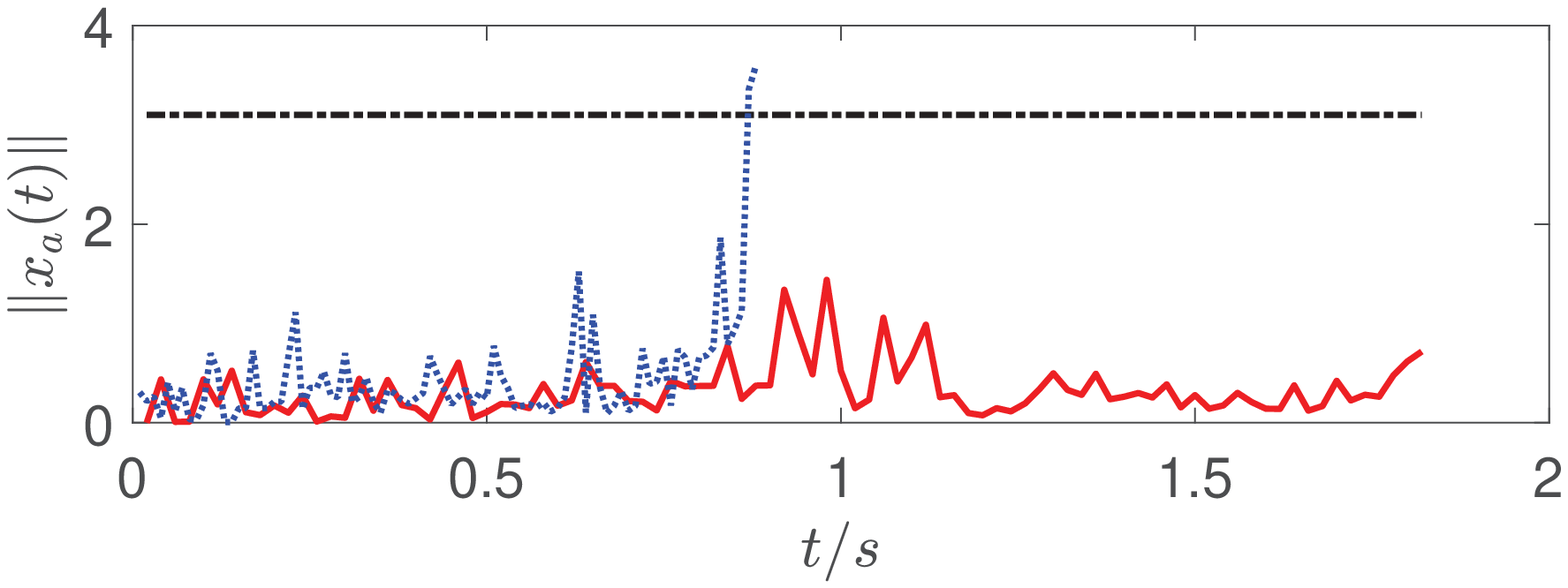}} \\
  \caption{Controlled output and detection results of NIPVSS under TPDAs \cite{JeonEun:19} with $\mathcal{A}_{c,t}(A_n)$ (\ref{eq3C1}) or the proposed MAPDAs with ${\mathcal{A}_{a,t}}({A_n},\Phi_n,x_a)$ (\ref{eq4B2}) of $Q=I$ and $Z=10000I$. Blue line: Under TPDAs with $\mathcal{A}_{c,t}(A_n)$ (\ref{eq3C1}). Red Line: Under MAPDAs with ${\mathcal{A}_{a,t}}({A_n},\Phi_n,x_a)$ (\ref{eq4B2}). Black line in (c): The  detection threshold $\epsilon=3.1$.}
  \label{fig6}
\end{figure}

The experiments are operated using the above set parameters, and the experimental results of NIPVSS under TPDAs with $\mathcal{A}_{c,t}(A_n)$ (\ref{eq3C1}) and the proposed MAPDAs with ${\mathcal{A}_{a,t}}({A_n},\Phi_n,x_a)$ (\ref{eq4B2}) are shown in Fig.~\ref{fig6}. When there exists model mismatch, we observe that: (1) TPDAs with $\mathcal{A}_{c,t}(A_n)$ (\ref{eq3C1}) drive the pendulum angle of NIPVSS to cross the maximum allowable angle 0.8 rad shown by the blue line in Fig.~\ref{fig6}(b), while the detector succeeds to detect them shown by the blue line in Fig.~\ref{fig6}(c), and (2) MAPDAs with ${\mathcal{A}_{a,t}}({A_n},\Phi_n,x_a)$ (\ref{eq4B2}) cannot be detected shown by the red line in Fig.~\ref{fig6}(c) before driving the cart position to cross the allowable limit -0.3 m shown by red line in Fig.~\ref{fig6}(a). It confirms that the proposed MAPDAs are not detectable before achieving successful destructiveness.

The above has presented rather good results for the proper parameters $Q$ and $Z$, however if the improper parameters $Q$ and $Z$ are chosen, it will produce less satisfactory experimental results of Figs. A.4 and A.5 in Section 4.2.2 of the supplementary materials \cite{DuZhang:22}.

\subsection{Performance of MAPDAs, DFLCAs and TLCAs}
For further comparison with the existing methods, three types of attacks are constructed: DFLCAs \cite{LiYang:18} using both the measurements and control input, TLCAs \cite{LiXie:19} using both the measurements and control input (or using only the measurements) and the proposed MAPDAs using only the measurements.

\begin{figure}[!t]
  \centering
  \subfigure[$\alpha(t)$]{\includegraphics[width=0.47\textwidth]{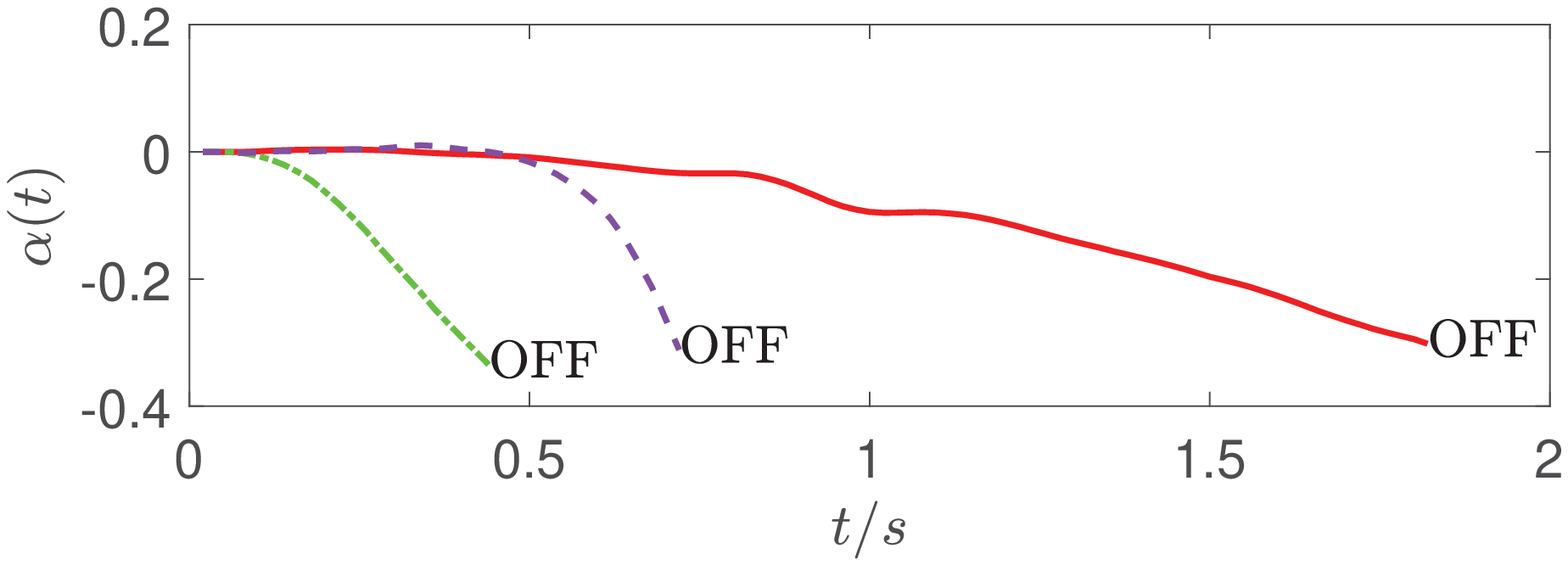}}\\ \vspace{-0.15in}
  \subfigure[$\theta(t)$]{\includegraphics[width=0.47\textwidth]{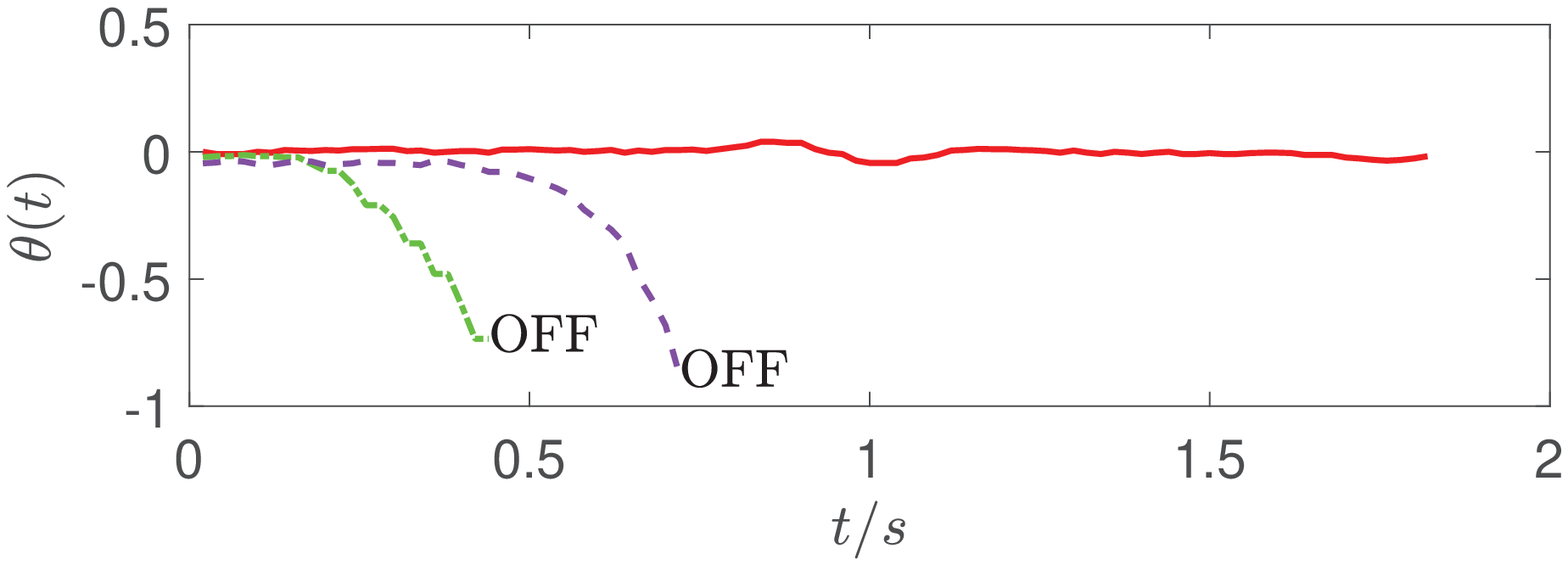}} \\ \vspace{-0.15in}
  \subfigure[$\left\| {x_a}(t) \right\|$]{\includegraphics[width=0.47\textwidth]{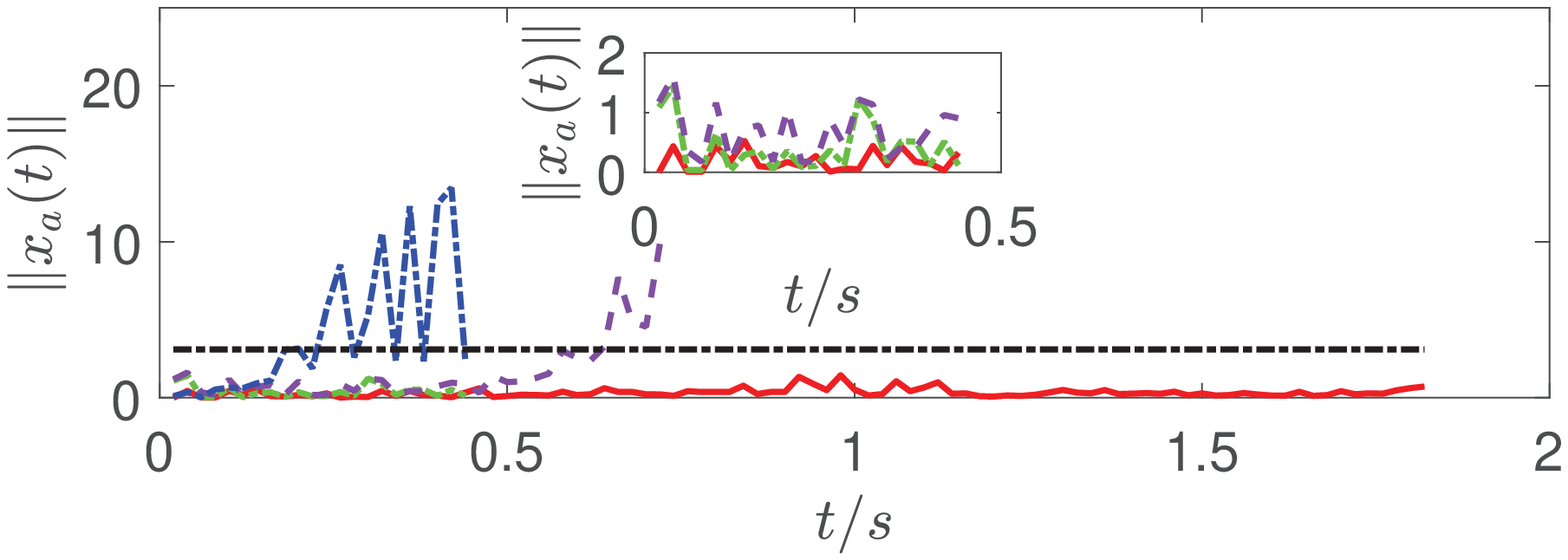}} \\
  \caption{Controlled output and detection results of NIPVSS under DFLCAs \cite{LiYang:18}, TLCAs \cite{LiXie:19} and the proposed MAPDAs with ${\mathcal{A}_{a,t}}({A_n},\Phi_n,x_a)$ (\ref{eq4B2}). Purple line: Under DFLCAs. Green line: Under TLCAs. Blue line in (c): Detection result of TLCAs without using control input. Red Line: Under MAPDAs with ${\mathcal{A}_{a,t}}({A_n},\Phi_n,x_a)$ (\ref{eq4B2}). Black line in (c): The detection threshold $\epsilon=3.1$.}
  \label{fig7}
\end{figure}

The experimental results of NIPVSS under DFLCAs, TLCAs and MAPDAs with ${\mathcal{A}_{a,t}}({A_n},\Phi_n,x_a)$ (\ref{eq4B2}) are shown in Fig.~\ref{fig7}. When there exists model mismatch, we observe that:  (1) DFLCAs drive the pendulum angle and cart position of NIPVSS to cross the allowable limit -0.8 rad and -0.3 m shown by the purple lines in Fig.~\ref{fig7}(a) and (b) respectively, while the detector succeeds to detect them shown by the purple line in Fig.~\ref{fig7}(c); (2) TLCAs drive the pendulum angle and cart position of NIPVSS to cross the allowable limit -0.8 rad and -0.3 m shown by the green lines in Fig.~\ref{fig7}(a) and (b) respectively, while the detector fails to detect them shown by the green line in Fig.~\ref{fig7}(c); it can be also seen from the blue line in Fig.~\ref{fig7}(c) that once TLCAs do not use control input (\emph{i.e.}, cannot construct the covert agent), they will be detected by the detector; (3) MAPDAs with ${\mathcal{A}_{a,t}}({A_n},\Phi_n,x_a)$ (\ref{eq4B2}) drive the cart position of NIPVSS to cross the allowable limit -0.3 m shown by the red line in Fig.~\ref{fig7}(a), while the detector fails to detect them shown by the red line in Fig.~\ref{fig7}(c). Therefore, compared with DFLCAs and TLCAs, the proposed MAPDAs method using only the measurements can bypass the detector, i.e., achieve successfully stealthy attack.

\section{Conclusion}
We have shown in this paper that stealthy attacks on vulnerable-sensor-network-only NCSs are possible, particularly when the measurements and adaptive control methods are employed. As a prototype, a stealthy MAPDAs method has been designed by using the measurements and adaptive control methods, and its promising performance on stealthiness and destructiveness are analysed by using convergence of measurements. In the future, the current work can be further extended to quasi-ideal attacks.

\begin{ack}                               
The work was supported in part by the National Science Foundation of China under Grant Nos. 92067106, 61773253, 61803252, and 61833011, 111 Project under Grant No. D18003, and Project of Science and Technology Commission of Shanghai Municipality under Grant Nos. 20JC1414000, 19500712300, 19510750300, 21190780300.  
\end{ack}



\end{document}


\begin{frontmatter}

\title{Supplementary Materials--Stealthy Measurement-Aided Pole-Dynamics Attacks with Nominal Models\thanksref{footnoteinfo}} 

\thanks[footnoteinfo]{This paper was not presented at any IFAC meeting. Corresponding author Changda Zhang. Email: changdazhang@shu.edu.cn.}

\author[SHU]{Dajun Du}\ead{ddj@i.shu.edu.cn},    
\author[SHU]{Changda Zhang}\ead{changdazhang@shu.edu.cn},               
\author[SHU]{Chen Peng}\ead{c.peng@shu.edu.cn},
\author[SHU]{Minrui Fei}\ead{mrfei@staff.shu.edu.cn},  
\author[UOL]{Huiyu Zhou}\ead{hz143@leicester.ac.uk}

\address[SHU]{Shanghai Key Laboratory of
Power Station Automation Technology, School of Mechatronic Engineering
and Automation, Shanghai University, Shanghai 200444, China}  
\address[UOL]{School of Computing and Mathematical Sciences, University of Leicester, Leicester
LE1 7RH, U.K.}             

\begin{abstract}                          
This is the supplementary document of the paper entitled ``Stealthy Measurement-Aided Pole-Dynamics Attacks with Nominal Models'' submitted to \emph{Automatica}. Section 1 gives the supplement on Introduction. Section 2 gives the supplement on Problem Formulation. Section 3 gives the supplement on Measurement-Aided Pole-Dynamics Attacks. Section 4 gives the supplement on Experimental Results and Discussion.
\end{abstract}

\end{frontmatter}

\section{Supplement on Introduction}
\subsection{Related Work}
In this section, we review the previous works on model-based attacks and their improved version. The unique features of our work are then discussed as well.

\emph{(1) Feedback-loop covert attacks and data-driven feedback-loop/two-loop covert attacks:} Feedback-loop covert attacks  [Mikhaylenko \& Zhang, 2021] requiring exact models are designed for arbitrary plants and inject different attack signals into the measurements and control inputs, where the dynamics of attack-free measurements are identical to that of the attacked measurements and thus the attack signals remain stealthy. Note that attack signals injected into control inputs push the controllable deviation from the previous state. Data-driven feedback-loop/two-loop covert attacks requiring no model can use the collected measurements and control input (1) to design attack signals injected into the measurements and control inputs based on subspace predictive control method  [Li \& Yang, 2018], or (2) to identify the controller parameters for adding attack signals and predicting measurements [Li \emph{et al.}, 2019] so that the expected state deviation can be achieved whilst the attacks remain stealthy.

\emph{(2) Zero-dynamics attacks and robust zero-dynamics attacks:} Zero-dynamics attacks [Teixeira \emph{et al.}, 2015] requiring exact model are designed for non-minimum phase plant and inject attack signals into the control input, where the unstable zero-dynamics of plant is duplicated and thus attack signals remain stealthy in output-nulling space. Robust zero-dynamics attacks [Park \emph{et al.}, 2019] requiring nominal model is designed for SISO system and can use the collected measurement and control input to design attack signals injected into the control input based on Byrnes-Isidori normal form and disturbance observer, so that the expected state deviation can be achieve and the attack remain stealthy.

\emph{(3) PDAs and robust PDAs:} PDAs [Kim \emph{et al.}, 2021] are designed for the plants with unstable pole dynamics and injecting attack signals into the measurements, where the dynamics of attack-free measurements is identical to those of the attacked measurements and thus attack signals remain stealthy. Robust PDAs [Jeon \& Eun, 2019] is designed for SISO systems and can use the collected control inputs to design attack signals injected into the measurements, based on the Byrnes-Isidori normal form and the disturbance observer so that the expected state deviation can be achieved whilst the attack remains stealthy.

In a summary, the drawbacks of the above work [Mikhaylenko \& Zhang, 2021], [Teixeira \emph{et al.}, 2015], [Kim \emph{et al.}, 2021], [Li \& Yang, 2018], [Li \emph{et al.}, 2019], [Park \emph{et al.}, 2019], [Jeon \& Eun, 2019] are listed as follows:
\begin{enumerate}
  \item
  The model-based attacks require exact models, unrealistic for many industry control systems.
  \item
  Apart from requiring nominal or no model, the improved model-based attacks require the measurements, control inputs or both.
  \item
  To the best of our knowledge, the existing improved model-based attacks cannot work without any control input.
\end{enumerate}
The above three drawbacks motivate our current study.

\subsection{Supplement on Notation}
Table~\ref{TabNota} summarizes the notations most frequently used throughout the paper.
\begin{table}[!t]
\centering
\caption{Table of Notations}
\label{TabNota}
\begin{tabular}{lll}
  \toprule
  $x(t)$                       & $\triangleq$ & System state \\
  $z(t)$                       & $\triangleq$ & Controlled output\\
  $\mathcal{A}_{c,t}(A)$       & $\triangleq$ & Exact auxiliary model\\
  $x_{\rm eam}(t)$             & $\triangleq$ & State of $\mathcal{A}_{c,t}(A)$ \\
  $x_{a}(t)$                   & $\triangleq$ & Input of Controller \\
  $\epsilon$                   & $\triangleq$ & Detection threshold \\
  $\mathcal{T}$                & $\triangleq$ & Time period of attack \\
  $\xi$                        & $\triangleq$ & Admissible state limit \\
  $\mathcal{A}_{c,t}(A_n)$     & $\triangleq$ & Nominal auxiliary model\\
  $x_{\rm nam}(t)$             & $\triangleq$ & State of $\mathcal{A}_{c,t}(A_n)$ \\
  $\mathcal{A}_{a,t}(A_n,\Phi,x_a)$,\\${\mathcal{A}_{a,t}}({A_n},\Phi_n,{x_a})$     & $\triangleq$ & Adaptive auxiliary model\\
  $x_{\rm aam}(t)$             & $\triangleq$ & State of $\mathcal{A}_{a,t}$ \\
  $F_a(t)$           & $\triangleq$ & Adaptive gain of $\mathcal{A}_{a,t}$\\
  \bottomrule
\end{tabular}
\end{table}

\section{Supplement on Problem Formulation}
\subsection{Lemma A.1 and Its Proof}
For exact auxiliary model ${\mathcal{A}_{c,t}}(A)$ (3), considering the eigenvalues $\lambda_i$ ($i=1,\ldots, p$) of the matrix $A$, $A$ can be expressed as $A = XJ{X^{ - 1}}$, where $X$ is a constant matrix and $J=diag\{J_1,\ldots,J_i,\ldots,J_{{r_1}}\}$, ${{r_1}}\leqslant p$ and when $\lambda_i$ is a single root, $J_i=\lambda_i$; when $\lambda_i$ is a $r_2$-fold root, \emph{i.e.}, $\lambda_i=\lambda_{i+1}=\ldots=\lambda_{i+{r_2}-1}$, then
\[{J_i} = \left[ {\begin{array}{*{20}{c}}
  {{\lambda _i}}&1&{} \\
  {}& \ddots &1 \\
  {}&{}&{{\lambda _{i + {{r_2}} - 1}}}
\end{array}} \right].\]
To conveniently present the following Lemma A.1, $\psi({t}): = {X^{ - 1}}x_{\rm eam}({t})$ is denoted.
\begin{lem}
If and only if $\psi({t_0})$ with $\left\|\psi({t_0})\right\|<\infty$ satisfies
\begin{enumerate}
  \item[i)]
  When $\lambda_i$ locates on the open right half-plane, there are two cases: If $\lambda_i$ is a single root, $\psi_{i}({t_0})=0$; If $\lambda_i$ is a ${r_2}$-fold root, $\psi_{i}({t_0})=\ldots=\psi_{i+{r_2}-1}({t_0})=0$;
  \item[ii)]
  When $\lambda_j$ locates on the closed left half-plane, if $\lambda_j$ is a ${r_2}$-fold root, $\psi_{j+1}({t_0})=\ldots=\psi_{j+{r_2}-1}({t_0})=0$.
\end{enumerate}
then the state $x_{\rm eam}(t)$ of exact auxiliary model ${\mathcal{A}_{c,t}}(A)$ (3) with unstable $A$ will converge to 0.
\end{lem}
\begin{pf}
Considering $A = XJ{X^{ - 1}}$, the solution of exact auxiliary model ${\mathcal{A}_{c,t}}(A)$ (3) is
\begin{equation}
{\psi}(t) = {e^{J(t - {t_0})}}\psi ({t_0}).
\label{eqLA11}
\end{equation}
Considering the eigenvalues $\lambda_i$ ($i=1,\ldots, p$) of matrix $A$, without loss of generality, we set that $\lambda_1$ is a ${r_2}$-fold root, \emph{i.e.}, $\lambda_1=\lambda_2=\ldots=\lambda_{{r_2}}$, and $\lambda_{{r_2}+1},\ldots,\lambda_p$ are all single roots. Then, the above (\ref{eqLA11}) can be re-written as
\begin{align}
&\label{eqLA12} {\psi}(t) = \hfill \\
&\left[ {\begin{array}{*{20}{c}}
  {{e^{{\lambda _1}t}}{\psi _1}({t_0}) + t{e^{{\lambda _1}t}}{\psi _2}({t_0}) +  \ldots  + \frac{{{t^{{{r_2}} - 1}}}}{{\left( {{{r_2}} - 1} \right)!}}{e^{{\lambda _1}t}}{\psi _{{{r_2}}}}({t_0})} \\
   \vdots  \\
  {{e^{{\lambda _1}t}}{\psi _{{r_2}-1}}({t_0}) + t{\psi _{{r_2}}}({t_0})} \\
  {{e^{{\lambda _1}t}}{\psi _{{r_2}}}({t_0})} \\
  {{e^{{\lambda _{{{r_2}} + 1}}t}}{\psi _{{{r_2}} + 1}}({t_0})} \\
   \vdots  \\
  {{e^{{\lambda _p}t}}{\psi _p}({t_0})}
\end{array}} \right]. \nonumber \hfill
\end{align}

\textbf{Sufficiency.} When the items (i) and (ii) of Lemma A.1 are satisfied, if the $r_2$-fold root $\lambda_1$ is located on the open right half-plane, $\psi_{1}({t_0})=\ldots=\psi_{{r_2}}({t_0})=0$; if $\lambda_1$ is located on the closed left half-plane, $\psi_{2}({t_0})=\ldots=\psi_{{r_2}}({t_0})=0$. It guarantees that $\psi_{1}(t),\ldots,\psi_{{r_2}}(t)$ will converge to 0. When the items (i) and (ii) of Lemma A.1 are satisfied, if the single root $\lambda_j$, $j \in \{{r_2}+1,\ldots,p\}$ is located on the open right half-plane, $\psi_{j}({t_0})=0$. It guarantee that $\psi_{{r_2}+1}(t),\ldots,\psi_{p}(t)$ will converge to 0. Therefore, the state $x_{eam}(t)$ of exact auxiliary model ${\mathcal{A}_{c,t}}(A)$ (3) with unstable $A$ will converge to 0.

\textbf{Necessity.} When the state $x_{eam}(t)$ of exact auxiliary model ${\mathcal{A}_{c,t}}(A)$ (3) with unstable $A$ converges to 0, it is necessary that $\psi_{2}({t_0})=\ldots=\psi_{{r_2}}({t_0})=0$. If the $r_2$-fold root $\lambda_1$ is located on the open right half-plane, it is also necessary that $\psi_{1}({t_0})=\ldots=\psi_{{r_2}}({t_0})=0$. Furthermore, if the single root $\lambda_j$, $j \in \{{r_2}+1,\ldots,p\}$ is located on the open right half-plane, it is necessary that $\psi_{j}({t_0})=0$. It completes the proof.
\end{pf}

\subsection{Proof of Theorem 1}
\begin{figure}[!t]
  \centering
  \subfigure[]{\includegraphics[width=0.4\textwidth]{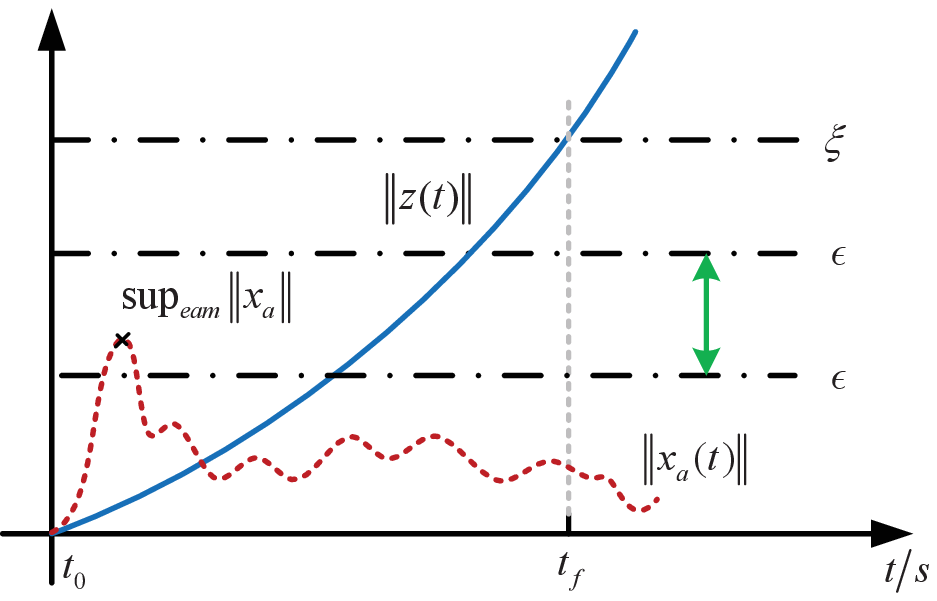}} \\ \vspace{-0.1in}
  \subfigure[]{\includegraphics[width=0.4\textwidth]{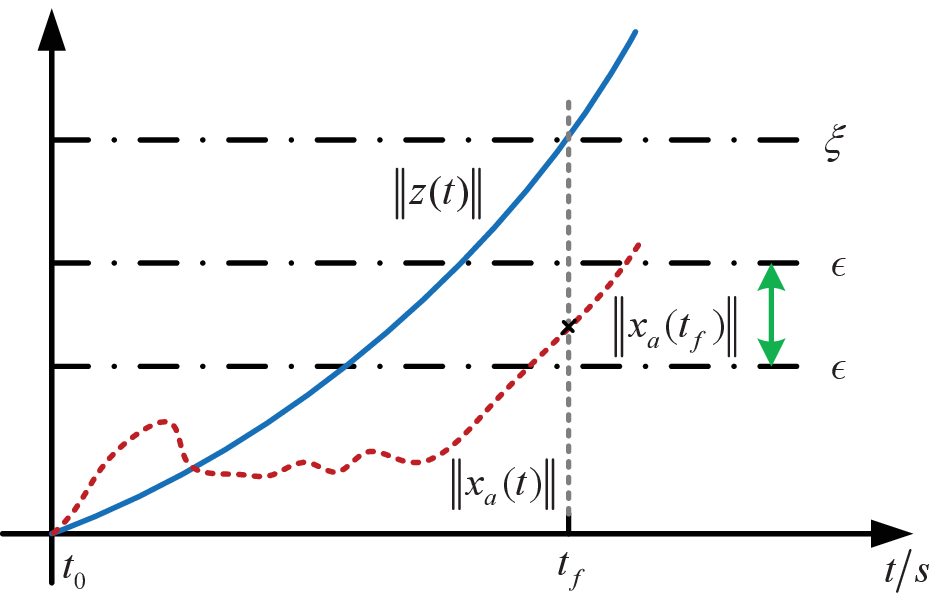}}
  \caption{(a) An illustration of states and detection results of NCSs under TPDAs with $\mathcal{A}_{c,t}(A)$ (3). (b) An illustration of states and detection results of NCSs under TPDAs with $\mathcal{A}_{c,t}(A_n)$ (10). Blue line: Values of $\left\| z(t) \right\|$. Red line: Values of $\left\| x_a(t) \right\|$. Green line: Change of $\epsilon$.}
  \label{figpam}
\end{figure}
Considering (1) and (3), the dynamics of $x_a(t)$ is
\begin{equation}
{\dot x_a}(t) = \Phi {x_a}(t).
\label{eqT11}
\end{equation}
If $\Phi$ is stable, it follows that
\begin{equation}
\left\| {{x_a}(t)} \right\| \leqslant {\sup}_{\rm eam} \left\| {x_a}\right\|, t \geqslant {t_0},
\label{eqT12}
\end{equation}
where ${\sup}_{\rm eam} \left\| {x_a}\right\| := \kappa \left\| {x_a(t_0)}\right\|$ and $\kappa$ is a constant. If $\epsilon > {\sup}_{\rm eam} \left\| {x_a}\right\|$, then (8) is guaranteed.

It can be seen from (\ref{eqT11}) and (3) that $\mathop {\lim }\limits_{t \to \infty } {x_a}(t) = 0$ if $\Phi$ is stable, and when $x_{\rm eam}(t_0)$ does not satisfy the item (i) or (ii) of Lemma A.1, $\mathop {\lim }\limits_{t \to \infty } \left\|{x_{\rm eam}}(t)\right\| \to \infty$ hold if $A$ is unstable. Considering (4), to make $\mathop {\lim }\limits_{t \to \infty } {x_a}(t) = 0$ hold, $\mathop {\lim }\limits_{t \to \infty } \left\|x(t)\right\| = 0$ must hold when $\mathop {\lim }\limits_{t \to \infty } \left\|{x_{\rm eam}}(t)\right\| \to \infty$ hold. Therefore, (9) is guaranteed. It completes the proof.

\subsection{Supplement on Remarks 4 and 6}
The illustration of the results of Theorems 1 and 2 are shown in Fig.~\ref{figpam}(a) and (b) respectively.

\subsection{Proof of Theorem 2}
Considering (1) and (10), the dynamics of $x_a(t)$ are
\begin{equation}
{\dot x_a}(t) = \Phi {x_a}(t) + (A - {A_n}){x_{\rm nam}}(t).
\label{eqT21}
\end{equation}
If $A_n$ is unstable, when $x_{\rm nam}(t_0)$ does not satisfy the item (i) or (ii) of Lemma A.2, $\mathop {\lim }\limits_{t \to \infty } \left\|x_{\rm nam}(t)\right\| \to \infty$ in (10) will hold. Therefore, although $\Phi$ is stable, (11) is guaranteed as $A \ne A_n$.

Considering (1), (4), (5) and (10), it follows that
\begin{equation}
\dot x(t) = \Phi x(t)-BK x_{\rm nam}(t).
\label{eqT22}
\end{equation}
Since $\mathop {\lim }\limits_{t \to \infty } \left\|x_{\rm nam}(t)\right\| \to \infty$ in (10) hold when $x_{\rm nam}(t_0)$ does not satisfy the item (i) or (ii) of Lemma A.2, the above (\ref{eqT22}) yields (9). It completes the proof.

\subsection{Lemma A.2 and Its Proof}
For nominal auxiliary model ${\mathcal{A}_{c,t}}(A_n)$ (10), considering the eigenvalues $\lambda_{n,i}$ ($i=1,\ldots, p$) of the matrix $A_n$, $A_n$ can be expressed as $A_n = X_nJ_n{X_n^{ - 1}}$, where $X_n$ is a constant matrix and $J_n=diag\{J_{n,1},\ldots,J_{n,i},\ldots,J_{n,{r_{n,1}}}\}$, ${r_{n,1}}\leqslant p$ and when $\lambda_{n,i}$ is a single root, $J_{n,i}=\lambda_{n,i}$; when $\lambda_{n,i}$ is a ${r_2}_n$-fold root, \emph{i.e.}, $\lambda_{n,i}=\lambda_{n,i+1}=\ldots=\lambda_{n,i+{r_{n,2}}-1}$, then
\[{J_{n,i}} = \left[ {\begin{array}{*{20}{c}}
  {{\lambda _{n,i}}}&1&{} \\
  {}& \ddots &1 \\
  {}&{}&{{\lambda _{n,i + {r_{n,2}} - 1}}}
\end{array}} \right].\]
To conveniently present the following Lemma A.2, $\psi_n({t}): = {X_n^{ - 1}}x_{\rm nam}({t})$ is denoted.
\begin{lem}
If and only if $\psi_n({t_0})$ with $\left\|\psi_n({t_0})\right\|<\infty$ satisfies
\begin{enumerate}
  \item[i)]
  When $\lambda_{n,i}$ locates on the open right half-plane, there are two cases: If $\lambda_{n,i}$ is a single root, $\psi_{n,i}({t_0})=0$; If $\lambda_{n,i}$ is a $r_{n,2}$-fold root, $\psi_{n,i}({t_0})=\ldots=\psi_{n,i+{r_{n,2}}-1}({t_0})=0$;
  \item[ii)]
  When $\lambda_{n,j}$ locates on the closed left half-plane, if $\lambda_{n,j}$ is a $r_{n,2}$-fold root, $\psi_{n,j+1}({t_0})=\ldots=\psi_{n,j+r_{n,2}-1}({t_0})=0$.
\end{enumerate}
then the state $x_{\rm nam}(t)$ of nominal auxiliary model ${\mathcal{A}_{c,t}}(A_n)$ (10) with unstable $A_n$ will converge to 0.
\end{lem}
\begin{pf}
The proof is similar as that of Lemma A.1, which is omitted.
\end{pf}

\section{Supplement on Measurement-Aided Pole-Dynamics Attacks}
\subsection{Proof of Theorem 3}
Considering the systems (1), (2), (4), (5) under MAPDAs with ${\mathcal{A}_{a,t}}({A_n},\Phi,x,x_a)$ (12), the dynamic of $x_a(t)$ is (13). Denoting $F_a^d(t) := F_a(t)+ {A_n}-A$, the dynamic of $x_a(t)$ (13) can be re-expressed as
\begin{equation}
\label{eqT31} {\dot x_a}(t) = \Phi {x_a}(t) - F_a^d(t){x_{\rm aam}}(t).
\end{equation}

To analyze the stability of $x_a(t)$, a candidate Lypunov function is chosen as
\begin{equation}
V(t) = x_a^T(t)P{x_a}(t) + tr\left( {{{\left( {F_a^d(t)} \right)}^T}Z^{ - 1}F_a^d(t)} \right).
\label{eqT32}
\end{equation}
Taking the derivative of $V(t)$ along with $t$ leads to
\begin{align}
  \dot V(t) &= x_a^T(t)\left( {{\Phi ^T}P + P\Phi } \right){x_a}(t) \nonumber \hfill \\
   &-2x_a^T(t)PF_a^d(t){x_{\rm aam}}(t) \nonumber \hfill \\
   &+ 2tr\left( {{{\left( {\dot F_a^d(t)} \right)}^T}Z^{ - 1}F_a^d(t)} \right). \hfill \label{eqT33}
\end{align}
If there are
\begin{equation}
\label{eqT34}
{\left( {\dot F_a^d(t)} \right)^T}Z^{ - 1} = {x_{\rm aam}}(t)x_a^T(t)P,
\end{equation}
and (12d), \emph{i.e.}, (12b) and (12d) hold true, and we have
\begin{equation}
\dot V(t) =  - x_a^T(t)Q{x_a}(t) < 0.
\label{eqT35}
\end{equation}
The above (\ref{eqT35}) yields (14). It completes the proof.

\subsection{Supplement on Remark 9}
\begin{figure}[!t]
  \centering
  \subfigure[]{\includegraphics[width=0.4\textwidth]{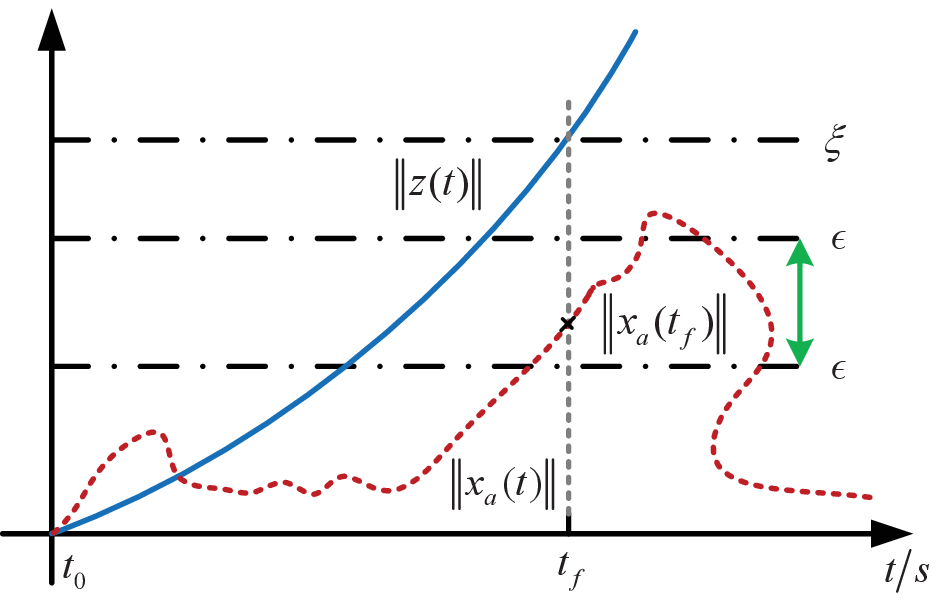}}\\ \vspace{-0.15in}
  \subfigure[]{\includegraphics[width=0.4\textwidth]{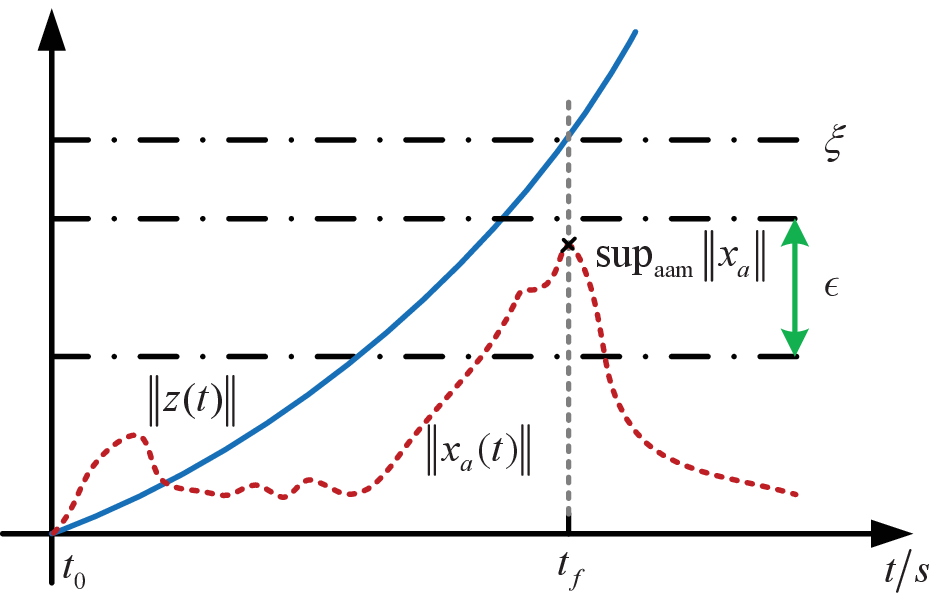}} \\ \vspace{-0.15in}
  \subfigure[]{\includegraphics[width=0.4\textwidth]{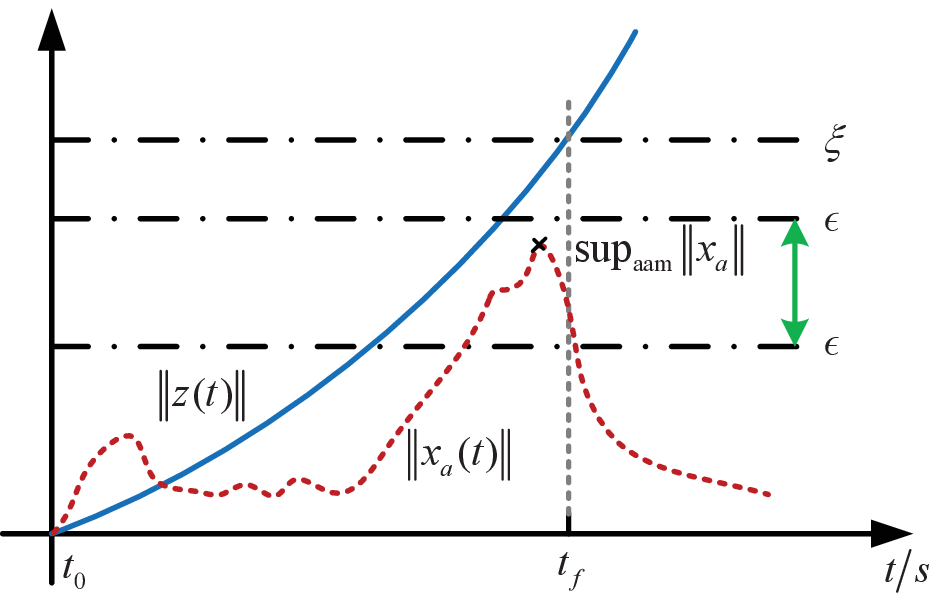}} \\
  \caption{An illumination of states and detection results of NCSs under MAPDAs with ${\mathcal{A}_{a,t}}({A_n},\Phi,x,{x_a})$ (15) or ${\mathcal{A}_{a,t}}({A_n},\Phi_n,x,x_a)$ (18). Blue line: Value of $\left\| x(t) \right\|$. Red line: Value of $\left\| x_a(t) \right\|$. Green line: Change of $\epsilon$. (a) Climbing Type. (b) Peak Type. (c) Descending Type.}
  \label{figaam}
\end{figure}
The illustration of the results of Theorem 3 is shown in Fig.~\ref{figaam}, where three types of MAPDAs are presented, i.e., climbing type ($\epsilon < {\sup _{\rm aam}}\left\| {{x_a}} \right\|$), peak type ($\epsilon = {\sup _{\rm aam}}\left\| {{x_a}} \right\|$) and descending type ($\epsilon > {\sup _{\rm aam}}\left\| {{x_a}} \right\|$).

\subsection{Supplement on Remark 12}
When the open-loop dynamic of the nominal system is stable (i.e., $A_n$ is stable), Eq. (14) in Theorem 3 will still hold. When (14) holds, $\left\|x_{\rm aam}(t)\right\|$ could be unbounded or convergent, i.e., $\left\|x(t)\right\|$ could be unbounded or convergent according to $x_a(t)=x(t)-x_{\rm aam}(t)$. The analysis is below using the contradiction method:
\begin{enumerate}
  \item[1)]
  Suppose that both $x_{\rm aam}(t)$ and $x_a(t)$ are convergent. Then, $\dot F_a(t)$ in (12b) is convergent, and thus $A-F_a(t)-A_n$ in (13) is bounded. Using (13), it can be obtained that $x_{a}(t)$ is convergent. Therefore, there is no contradiction with (14).
  \item[2)]
  Suppose that $\left\|x_{\rm aam}(t)\right\|$ is unbounded and $x_a(t)$ is convergent. Then, $\dot F_a(t)$ in (12b) could be convergent because $x_a(t)x^{T}_{\rm aam}(t)$ is undetermined type, and thus $A-F_a(t)-A_n$ in (13) could be convergent. Using (13), we understand that $x_{a}(t)$ are convergent. Therefore, there is no contradiction with (14).
\end{enumerate}
Therefore, when the open-loop dynamic of the nominal system is stable, the proposed MAPDAs cannot ensure that the system state is divergent.

\subsection{Supplement on Remark 13}
The attacker can adopt discrete-time TPDAs and MAPDAs transformed from continuous TPDAs (3) or (10) and MAPDAs (12) as follows:
\begin{enumerate}
  \item[1)]
  TPDAs can use a discrete-time exact auxiliary model
  \begin{subequations}
  \label{eqR121}
  \begin{align}
  \label{eqR121a}  {x_{\rm eam}}\left( {(k+1)h} \right) &= A^h{x_{\rm eam}}(kh), \hfill \\
  \label{eqR121b}  a(kh) &= {x_{\rm eam}}(kh), \hfill
  \end{align}
  \end{subequations}
  where $k=0,1,2,\ldots$, ${A^h} := e^{Ah}$ and $h$ is the sampling period.
  \item[2)]
  TPDAs can use a discrete-time nominal auxiliary model
  \begin{subequations}
  \label{eqR122}
  \begin{align}
  \label{eqR122a}  {x_{\rm nam}}\left( {(k+1)h} \right) &= A_n^h{x_{\rm nam}}(kh), \hfill \\
  \label{eqR122b}  a(kh) &= {x_{\rm nam}}(kh), \hfill
  \end{align}
  \end{subequations}
  where $A_n^{h} := e^{A_n h}$.
  \item[3)]
  MAPDAs can use a discrete-time adaptive auxiliary model
  \begin{subequations}
  \label{eqR123}
  \begin{align}
  \label{eqR123a}{x_{{\text{aam}}}}\left( {(k + 1)h} \right) &= \Xi_n(kh){x_{{\text{aam}}}}(kh), \hfill \\
  {F_a}\left( {(k + 1)h} \right) &= {F_a}(kh)  \nonumber \hfill \\
  \label{eqR123b}&+ hZP{x_a}(kh)x_{{\text{aam}}}^T(kh), \hfill \\
  \label{eqR123c} a(kh) &= {x_{{\text{aam}}}}(kh), \hfill \\
  \label{eqR123d}- Q &= \Phi _n^TP + P{\Phi _n}, \hfill
  \end{align}
  \end{subequations}
  where ${\Xi _n}(kh) := {e^{\left( {{A_n} + {F_a}(kh)} \right)h}}$.
\end{enumerate}

The next aim is to analyse the effectiveness of the above discrete-time MAPDAs (\ref{eqR123}) (taken as an example) for the digital system. Considering that when the sensors and controller are digitized, the digital system becomes a sample-data-based hybrid system and can be expressed as the following time-delay model:
\begin{subequations}
\label{eqR124}
\begin{align}
  \label{eqR124a}\dot x(t) &= Ax(t) + Bu(t), [\emph{i.e.}, (1)], \hfill \\
  \label{eqR124b}{x_a}(t) &= x(t) - a(t), [\emph{i.e.}, (4)], \hfill \\
  \label{eqR124e}u(t) &= K{x_a}(t - {\tau _t}), \hfill
\end{align}
\end{subequations}
where $t \in [kh,(k+1)h)$, $k=0,1,2,\ldots$, and $\tau_t \in [0,h)$ is time delay and $\dot \tau_{t}=1$. Based on the above time-delay model (\ref{eqR124}), the stability criterion on delay-induced continuous MAPDAs to guarantee (14) will be presented in the following Theorem A.1.
\setcounter{thm}{0}
\begin{thm}
Considering the constructed time-delay model (\ref{eqR124}), for a scalar $h$, the matrices $P_1>0$, $P_2>0$, $P_3>0$, $P_4>0$ and $Z_1>0$, if there are delay-induced continuous MAPDAs
\begin{subequations}
\label{eqR125}
\begin{align}
\label{eqR125a}{\dot x_{\rm aam}}(t) &= \left({A_n}+F_a(t)\right){x_{\rm aam}}(t), \hfill \\
\label{eqR125b}\dot F_a(t ) &= Z_1{P_1}{x_a}(t)x_{\text{aam}}^T(t) \hfill \\
&+ {h^2}Z_1{P_4}\left( {A{x_a}(t) + BK{x_a}(t - {\tau _t})} \right)x_{\text{aam}}^T(t) \nonumber \hfill \\
&- \frac{1}{2}{h^2}Z_1{P_4}F_a^d(t){x_{{\text{aam}}}}(t)x_{\text{aam}}^T(t), \nonumber \hfill \\
\label{eqR125c}a(t) &= {x_{\rm aam}}(t), \hfill \\
\label{eqR125d}- Q &= {\Phi ^T}P + P\Phi, \hfill
\end{align}
\end{subequations}
$F_a^d(t)$ have been given in (\ref{eqT31}), and
\begin{equation}
\Omega  = \left[ {\begin{array}{*{20}{c}}
  {{\Omega _{11}}}&{{\Omega _{12}}}&{{\Omega _{13}}} \\
   * &{{\Omega _{22}}}&{{\Omega _{23}}} \\
   * & * &{{\Omega _{33}}}
\end{array}} \right]<0,
\label{eqR126}
\end{equation}
where ${\Omega _{11}} = {A^T}{P_1} + {P_1}A + {P_2} + {P_3} + {h^2}{A^T}{P_4}A - {P_4}$, ${\Omega _{22}} =  - {P_2} + {h^2}{K^T}{B^T}{P_4}BK$, ${\Omega _{33}} =  - {P_3} - {P_4}$, ${\Omega _{12}} = {P_1}BK + {h^2}{A^T}{P_4}BK$, ${\Omega _{13}} = {P_4}$, and ${\Omega _{23}} = 0$, then (14) will hold.
\end{thm}
\begin{pf}
From the above (\ref{eqR124}), (\ref{eqR125a}) and (\ref{eqR125c}), the dynamics of $x_a(t)$ becomes
\begin{equation}
{{\dot x}_a}(t) = A{x_a}(t) +BK{x_a}(t-\tau_t)- {F^d_a}(t) {x_{{\text{aam}}}}(t),
\label{eqR127}
\end{equation}

To analyze the stability of $x_a(t)$, we choose the candidate Lypunov function
\begin{align}
  {V_{tds}}(t) &= x_a^T(t)P_1{x_a}(t) + \int_{t - {\tau _t}}^t {x_a^T(s){P_2}{x_a}(s)ds}  \nonumber \hfill \\
   &+ \int_{t - h}^t {x_a^T(s){P_3}{x_a}(s)ds}  \nonumber \hfill \\
   &+ \int_{ - h}^0 {\int_{t + v}^t {h\dot x_a^T(s){P_4}{{\dot x}_a}(s)ds} dv}  \nonumber \hfill \\
   &+ tr\left( {{{\left( {F_a^d(t)} \right)}^T}Z_1^{ - 1}F_a^d(t)} \right). \hfill \label{eqR128}
\end{align}

Taking the derivative of $V_{tds}(t)$ along with $t$, denoting ${\chi _a}(t) = \left[ {{x_a}(t);{x_a}(t - {\tau _t});{x_a}(t - h)} \right]$ and using Jensen's inequality \cite{Han:05} leads to
\begin{align}
  &{{\dot V}_{tds}}(t) \leqslant \chi _a^T(t)\Omega {\chi _a}(t) \label{eqR129} \hfill \\
   &+ tr\left( {{{\left( {F_a^d(t)} \right)}^T}Z_1^{ - 1}\dot F_a^d(t)} \right) \nonumber \hfill \\
   &- {\left( {F_a^d(t){x_{{\text{aam}}}}(t)} \right)^T}{P_1}{x_a}(t) \nonumber \hfill \\
   &- {h^2}{\left( {F_a^d(t){x_{{\text{aam}}}}(t)} \right)^T}{P_4}\left( {A{x_a}(t) + BK{x_a}(t - {\tau _t})} \right) \nonumber \hfill \\
   &+ \frac{1}{2}{h^2}{\left( {F_a^d(t ){x_{{\text{aam}}}}(t)} \right)^T}{P_4}F_a^d(t){x_{{\text{aam}}}}(t) \nonumber \hfill \\
   &+ tr\left( {{{\left( {\dot F_a^d(t )} \right)}^T}Z_1^{ - 1}F_a^d(t)} \right) \nonumber \hfill \\
   &- x_a^T(t){P_1}F_a^d(t){x_{{\text{aam}}}}(t) \nonumber \hfill \\
   &- {h^2}{\left( {A{x_a}(t) + BK{x_a}(t - {\tau _t})} \right)^T}{P_4}F_a^d(t ){x_{{\text{aam}}}}(t) \nonumber \hfill \\
   &+ \frac{1}{2}{h^2}{\left( {F_a^d(t){x_{{\text{aam}}}}(t)} \right)^T}{P_4}F_a^d(t){x_{{\text{aam}}}}(t). \nonumber \hfill
\end{align}

If there are (\ref{eqR125}) and (\ref{eqR126}), then
\begin{equation}
\dot V_{tds}(t) \leqslant  x_a^T(t)\Omega{x_a}(t) < 0.
\label{eqR1210}
\end{equation}
The above (\ref{eqR1210}) yields (14). It completes the proof.
\end{pf}
According to (\ref{eqR125}) in Theorem A.1, a delay-induced discrete-time MAPDAs can be obtained as
\begin{subequations}
\label{eqR1211}
\begin{align}
\label{eqR1211a}{x_{{\text{aam}}}}\left( {(k + 1)h} \right) &= \Xi_n(kh){x_{{\text{aam}}}}(kh), \hfill \\
\label{eqR1211b}{F_a}\left( {(k + 1)h} \right) &= {F_a}(kh)  \hfill \\
&+ hZ_1P_1{x_a}(kh)x_{{\text{aam}}}^T(kh) \nonumber \hfill \\
&+ {h^2}Z_1{P_4} A{x_a}(kh) x_{\text{aam}}^T(kh) \nonumber \hfill \\
&+ {h^2}Z_1{P_4} BK{x_a}((k-1)h)x_{\text{aam}}^T(kh) \nonumber \hfill \\
&- \frac{1}{2}{h^2}Z_1{P_4}F_a^d(kh){x_{{\text{aam}}}}(kh)x_{\text{aam}}^T(kh), \nonumber \hfill \\
\label{eqR1211c}a(kh) &= {x_{{\text{aam}}}}(kh), \hfill \\
\label{eqR1211d}- Q &= \Phi _n^TP + P{\Phi _n}, \hfill
\end{align}
\end{subequations}
When the sampling period is small, discrete-time MAPDAs (\ref{eqR123}) is a proper approximation of delay-induced discrete-time MPADAs (\ref{eqR1211}). Since delay-induced discrete-time MPADAs (\ref{eqR1211}) can guarantee the stealthiness (14), discrete-time MAPDAs (\ref{eqR123}) will be effective on guaranteeing the stealthiness (14) for the digital system.

\section{Supplement on Experimental Results and Discussion}
\subsection{Supplement on Threshold of the Detector}
\begin{figure}[!t]
  \centering
  \subfigure[Cart position]{\includegraphics[width=0.47\textwidth]{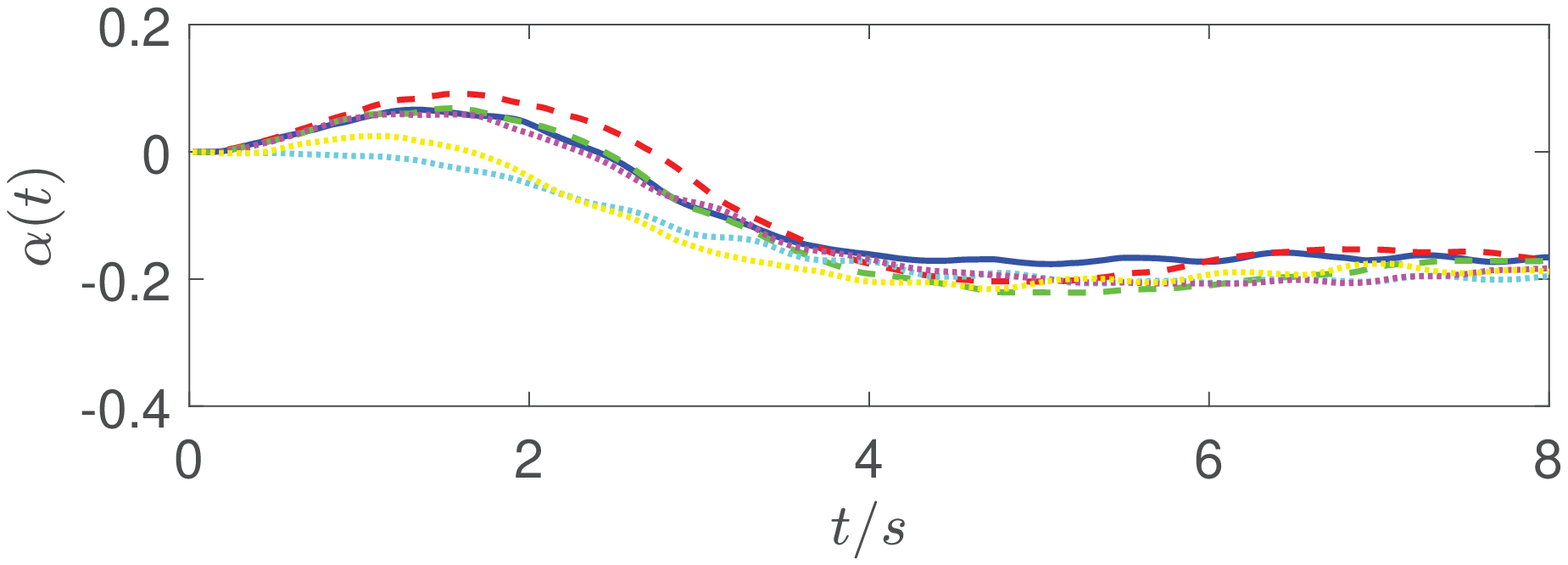}}  \\ \vspace{-0.15in}
  \subfigure[Pendulum Angle]{\includegraphics[width=0.47\textwidth]{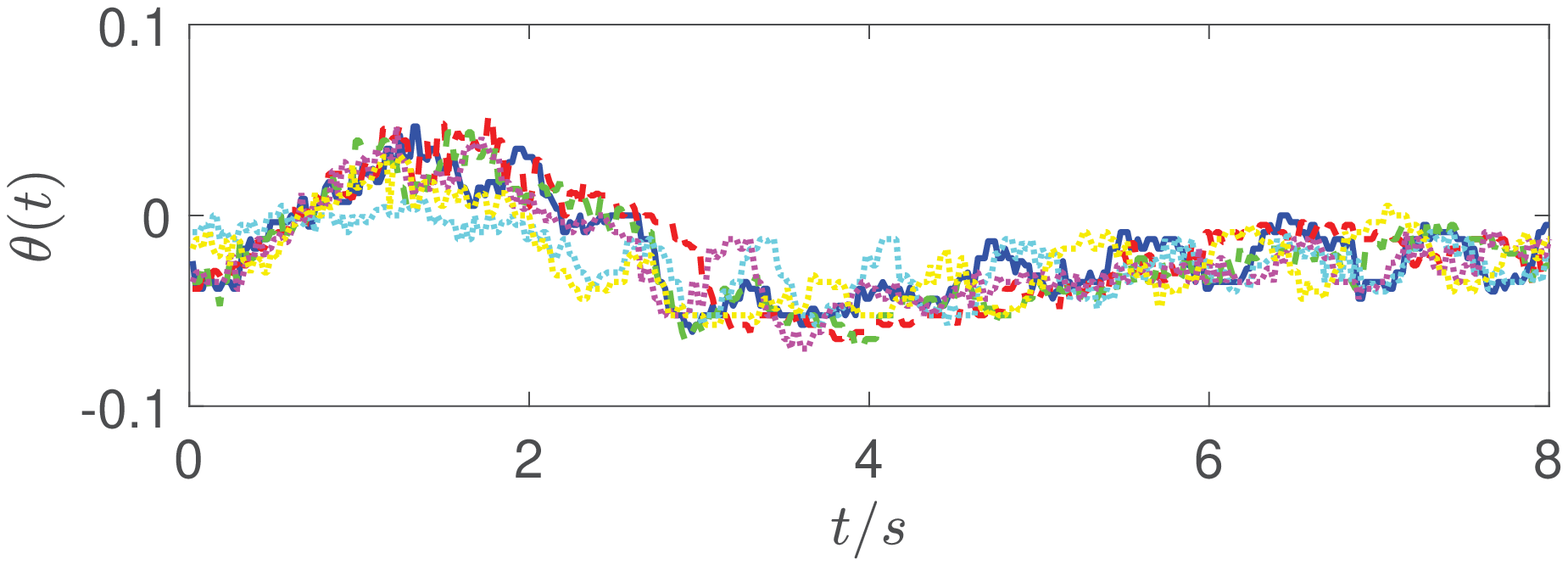}} \\ \vspace{-0.15in}
  \subfigure[Cart velocity]{\includegraphics[width=0.47\textwidth]{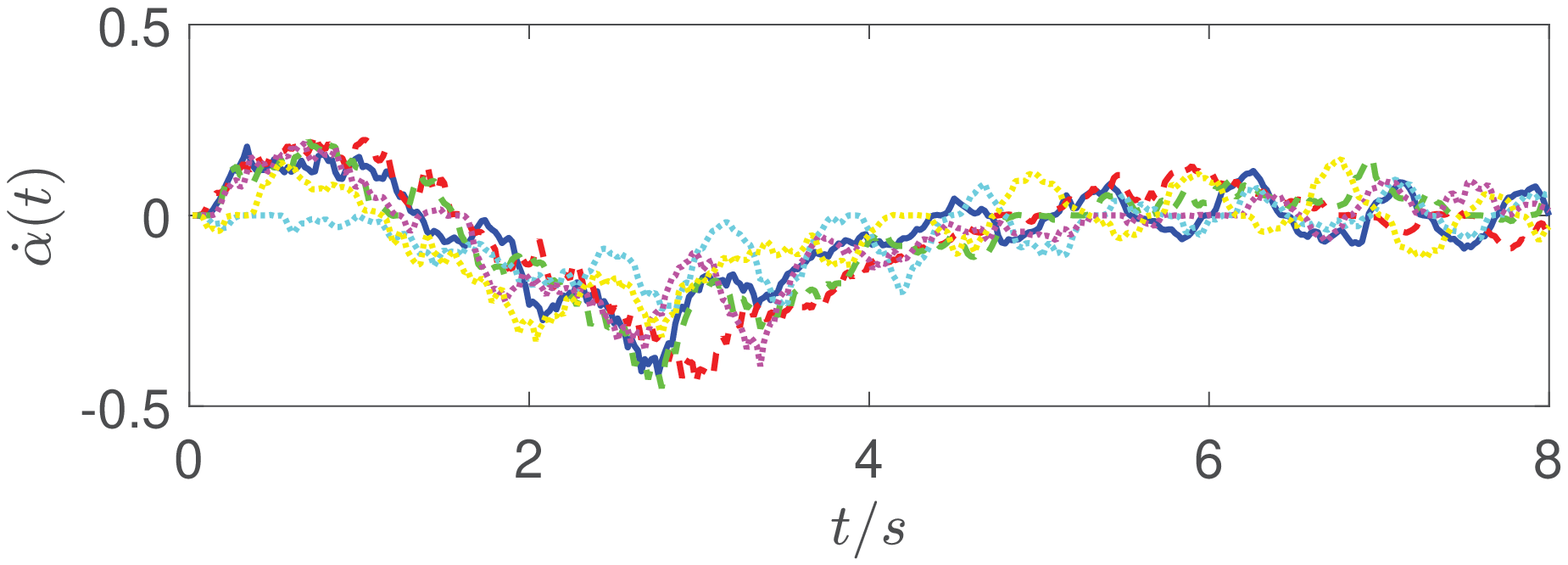}} \\ \vspace{-0.15in}
  \subfigure[Angular velocity]{\includegraphics[width=0.47\textwidth]{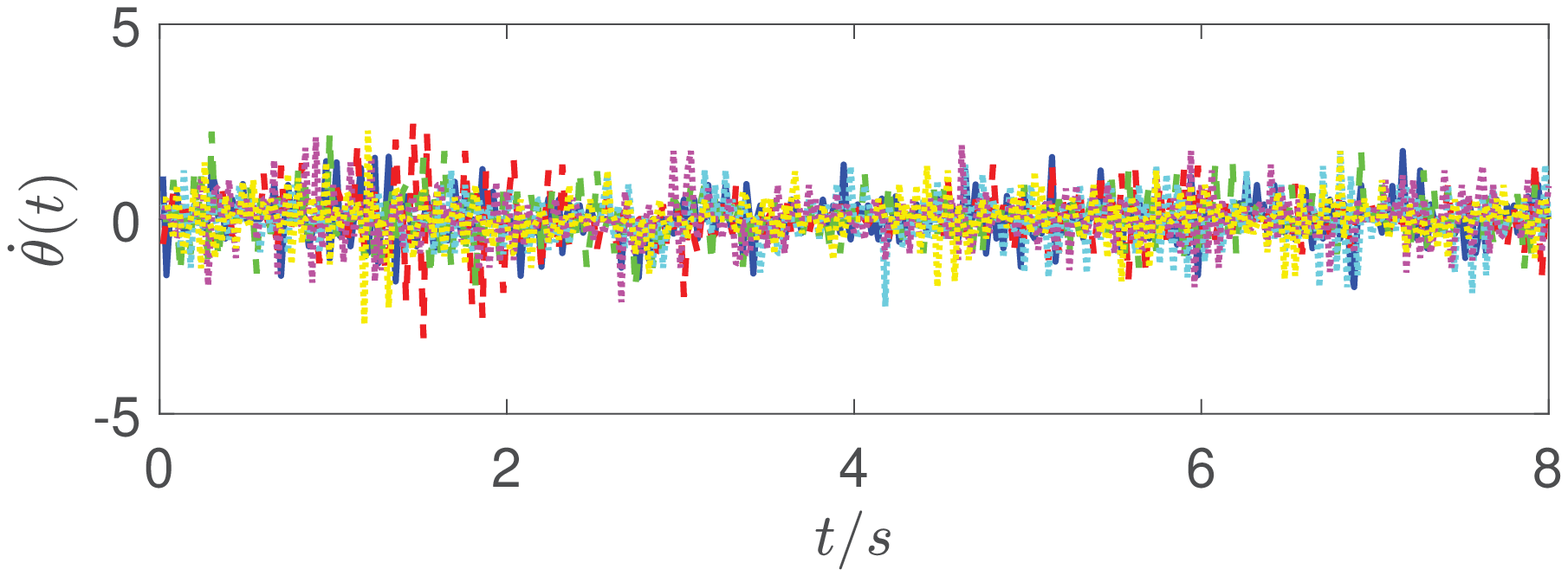}} \\ \vspace{-0.15in}
  \subfigure[Detection Results]{\includegraphics[width=0.47\textwidth]{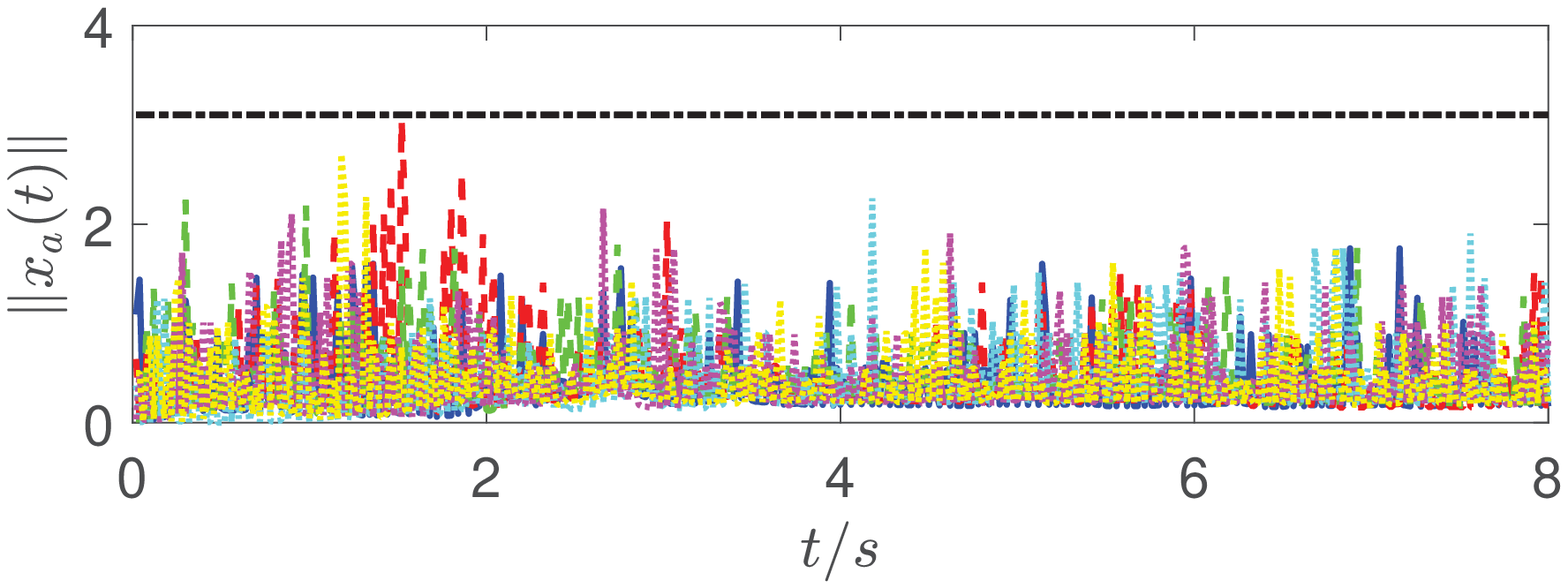}} \\
  \caption{States and detection results of attack-free NIPVSS. Blue, red, green, cyan, magenta and yellow lines: Results of different experiments. Black line in (e): The detection threshold is set as $\epsilon=3.1$.}
  \label{fig5}
\end{figure}
Due to page limitation, only six attack-free experimental results of NIPVSS are shown in Fig.~\ref{fig5}. The values of ${\sup^e_f}\left\| {{x_a}} \right\|$ for five experiments are listed in Table~\ref{Tabthd}.
\begin{table}[!t]
\centering
\caption{The values of $\sup^e_{f} \left\|{x_a}\right\|$ for Six Attack-Free Experiments}
\label{Tabthd}
\begin{tabular}{cccc}
\toprule
  ~  & $\sup^e_{f} \left\|{x_a}\right\|$ & ~ & $\sup^e_{f} \left\|{x_a}\right\|$ \\
\midrule
  Experiment 1 & 2.2544 & Experiment 4 & 2.2607 \\
  Experiment 2 & 3.0560 & Experiment 5 & 2.5429 \\
  Experiment 3 & 2.2544 & Experiment 6 & 2.7057 \\
\bottomrule
\end{tabular}
\end{table}

\subsection{Supplement on Performance of TPDAs and MAPDAs}
\subsubsection{Verification of $x_{\rm nam}(0)=0.0001 \textbf{1}$ not satisfying Lemma A.2}
Considering $A_n=X_nJ_nX_n^{-1}$ where
\[{X_n} = \left[ {\begin{array}{*{20}{c}}
  1&0&0&0 \\
  0&0&{0.5}&{0.5} \\
  0&1&0&0 \\
  0&0&{ - 2.7125}&{2.7125}
\end{array}} \right],
  {J_n} = \left[ {\begin{array}{*{20}{c}}
  0&1&0&0 \\
  0&0&0&0 \\
  0&0&{ - 5.425}&0 \\
  0&0&0&{5.425}
\end{array}} \right],\]
the eigenvalues of $J_n$ is $\lambda_{n,1}=\lambda_{n,2}=0$ (2-fold root), $\lambda_{n,3}=-5.425$ (single root), and $\lambda_{n,4}=5.425$ (single root). Considering $\psi_n(0) = {X_n^{ - 1}}{x_{\rm nam}}(0)=[0.1;0.1;0.0816;0.1184]\times 10^{-3}$, all the elements of $\psi_n(0)$ are not zero, which does not satisfy the item (i) or (ii) of Lemma A.2. Therefore, it satisfies the initial condition of Theorem 2.

\subsubsection{Performance of TPDAs and MAPDAs with different $Q$ and $Z$}
\begin{figure}[!t]
  \centering
  \subfigure[$\alpha(t)$]{\includegraphics[width=0.47\textwidth]{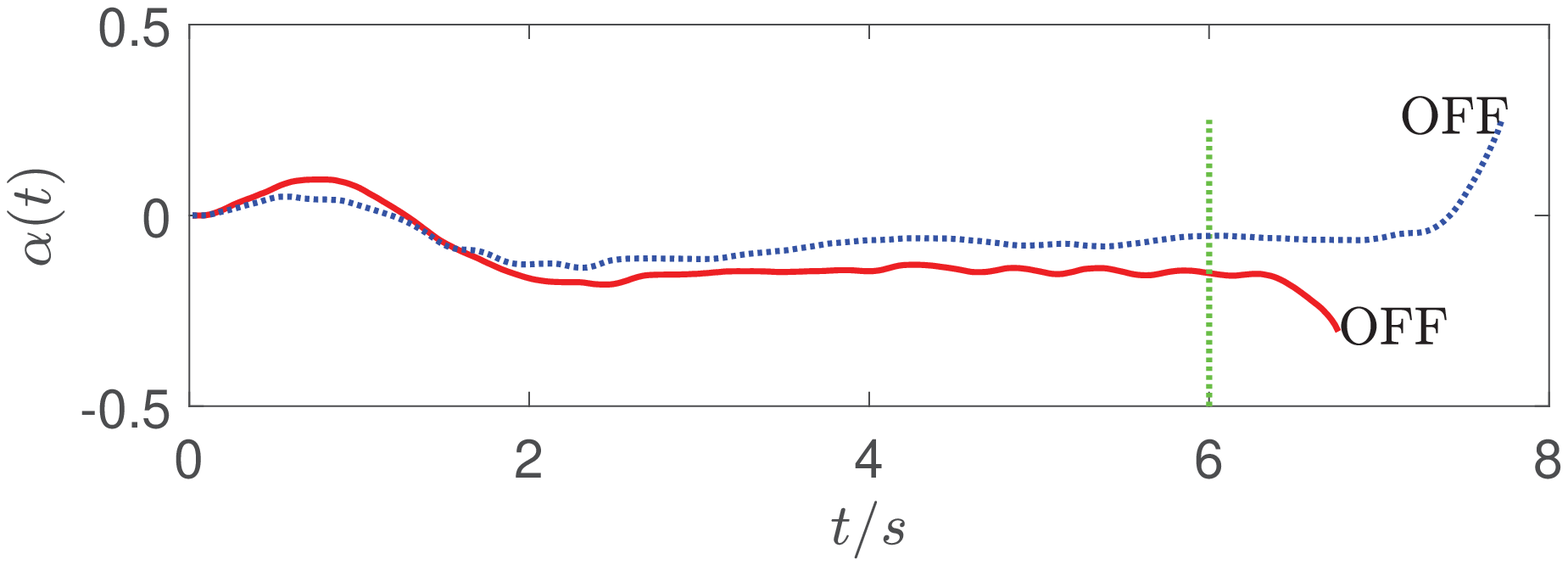}}\\ \vspace{-0.15in}
  \subfigure[$\theta(t)$]{\includegraphics[width=0.47\textwidth]{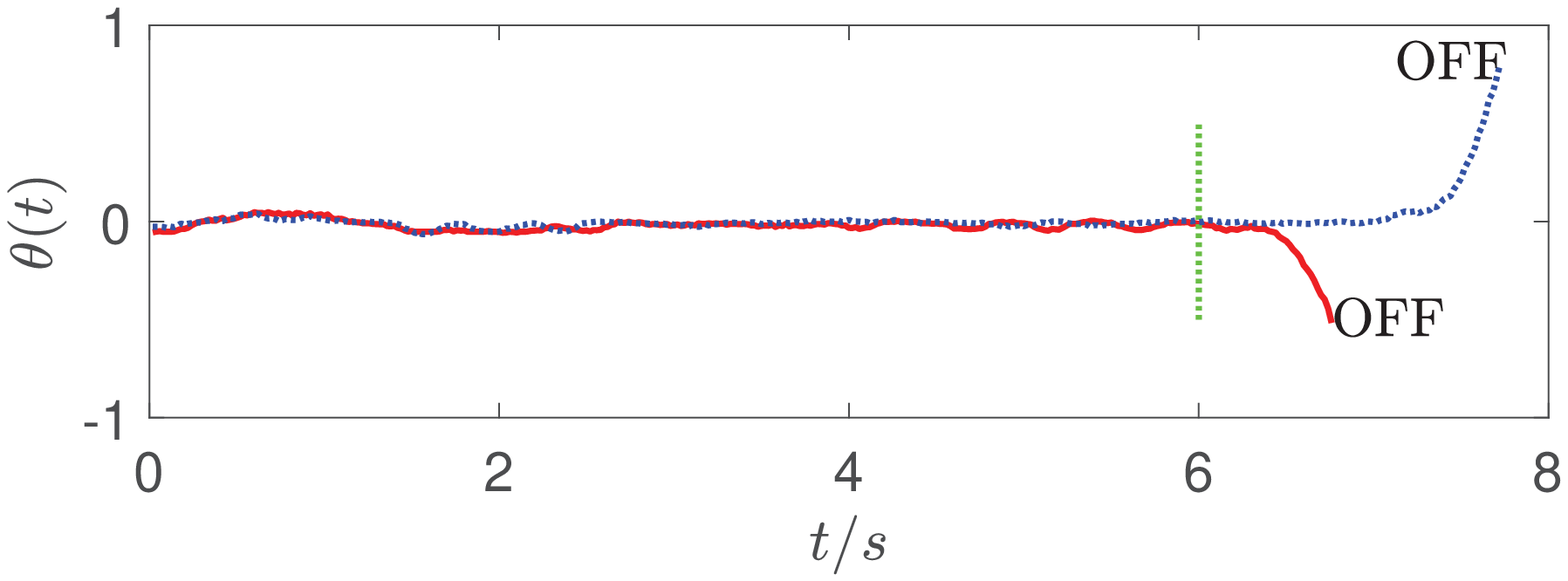}} \\ \vspace{-0.15in}
  \subfigure[$\left\| {x_a}(t) \right\|$]{\includegraphics[width=0.47\textwidth]{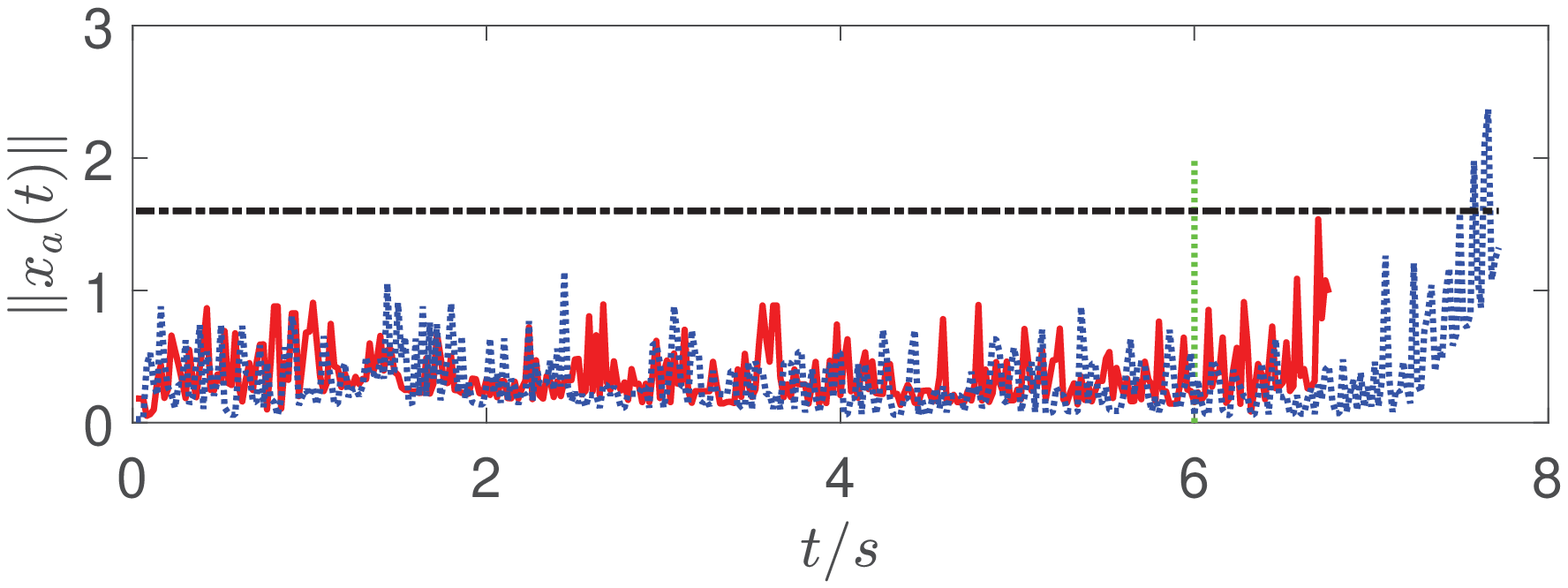}} \\
  \caption{Controlled output and detection results of NIPVSS under TPDAs  with $\mathcal{A}_{c,t}(A_n)$ (10) or the proposed MAPDAs with ${\mathcal{A}_{a,t}}({A_n},\Phi_n,x_a)$ (15) of $Q=I$ and $Z=0.5I$. Blue line: Under TPDAs with $\mathcal{A}_{c,t}(A_n)$ (10). Red Line: Under MAPDAs with ${\mathcal{A}_{a,t}}({A_n},\Phi_n,x_a)$ (15). Green line: Instant when attacks start. Black line in (c): The  detection threshold $\epsilon=1.6$.}
  \label{fig6o}
\end{figure}
\begin{figure}[!t]
  \centering
  \includegraphics[width=0.4\textwidth]{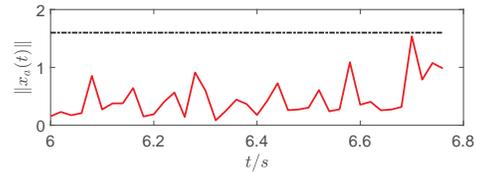}
  \caption{Detection results when $t \geqslant 6s$ of Fig.~\ref{fig6o}(c). Red Line: Under MAPDAs with ${\mathcal{A}_{a,t}}({A_n},\Phi_n,x,x_a)$ (15). Black line: The set detection threshold $\epsilon=1.6$.}
  \label{fig7o}
\end{figure}
The proper parameters $Q$ and $Z$ have produced rather good results as shown in the main paper, however if the improper parameters $Q$ and $Z$ are chosen, it will produce unsatisfactory experimental results. For example, when $Q=I$ and $Z=0.5I$, Figs.~\ref{fig6o} and~\ref{fig7o} show the experiments of NIPVSS under TPDAs with $\mathcal{A}_{c,t}(A_n)$ (10) when $t\geqslant 6s$ or the proposed MAPDAs with ${\mathcal{A}_{a,t}}({A_n},\Phi_n,x_a)$ (15) when $t\geqslant 6s$. It can be seen from Figs. A.4 and A.5 that the upper bound of $\left\| x_a(t) \right\|$ will almost touch the threshold ($\epsilon=1.6$ from 5 attack-free experiments) and the limit-crossing speed of $\left\| x(t) \right\|$ is high (as shown by the result that the blue lines touch the threshold later than red lines). It indicates that the parameters $Q$ and $Z$ are improperly chosen.